\documentstyle[11pt,aaspp4]{article}
\input tex.def
\slugcomment{To appear in {\it The Astrophysical Journal}.}
\lefthead{Ho}
\righthead{Spectral Energy Distributions}

\begin{document}

\title{The Spectral Energy Distributions of Low-Luminosity Active 
Galactic Nuclei}

\author{Luis C. Ho\footnote{Current address: Carnegie Observatories,
813 Santa Barbara St., Pasadena, CA 91101-1292.}}
\affil{Harvard-Smithsonian Center for Astrophysics, 60 Garden St., Cambridge, 
MA 02138}

\begin{abstract}
As a step toward elucidating the physical conditions in nearby active 
galaxies, this paper presents spectral energy distributions (SEDs) of a sample 
of seven low-luminosity active galactic nuclei (AGNs).  
SEDs for four objects are presented for the first time (NGC 4261, NGC 4579, 
NGC 6251, and M84); the data for the remaining three (M81, M87, and NGC 4594) 
have been substantially updated compared to previous studies.  The nuclear 
fluxes were carefully selected so as to avoid contamination by emission from 
the host galaxy, which can be substantial for very weak nuclei.  

The present sample of low-luminosity nuclei all exhibit SEDs that look 
markedly different from the canonical broad-band continuum spectrum of 
luminous AGNs.  The most striking difference is that the 
low-luminosity objects lack an ultraviolet excess (the ``big blue bump''), a 
feature normally associated with emission from a standard optically thick, 
geometrically thin accretion disk.  The weakness of the ultraviolet band leads 
to an unusually steep optical-ultraviolet continuum shape and a more pronounced 
contribution from the X-rays to the ionizing spectrum.  It is argued that 
the absence of the big blue bump is a property intrinsic to the SEDs and 
not an artifact of strong dust extinction.  Another notable 
property of the SEDs is the prominence of the compact, flat-spectrum radio 
component relative to the emission in other energy bands.  All seven nuclei in 
the sample, including three hosted by spiral galaxies, technically qualify as 
``radio-loud'' objects according to conventional criteria.  Finally, the 
integrated spectra confirm the exceptional weakness of the nuclei: 
the bolometric luminosities range from 2\e{41} to 8\e{42} \lum, or 
$\sim 10^{-6}-10^{-3}$ times the Eddington rate for the black hole masses 
previously reported for these galaxies.

\end{abstract}


\section{Introduction}

The spectral energy distributions (SEDs) of active galactic nuclei (AGNs) 
carry important information on the physical processes of the accretion 
process.  Many aspects of the AGN phenomenon, including the SED, 
have been successfully interpreted within the accretion disk framework, 
specifically one in which the disk is assumed to be optically thick and 
physically thin (Blandford \& Rees 1992, and references therein).  Previous 
work, however, has concentrated nearly exclusively on high-luminosity AGNs --- 
mainly bright Seyfert nuclei and QSOs.  Very little data exist on the spectral 
properties of low-luminosity AGNs, such as those commonly found in nearby 
galaxies (Ho, Filippenko, \& Sargent 1997a), because they are 
difficult to study.  Yet, knowledge of the SEDs of AGNs in the low-luminosity 
regime is fundamental to understanding the physical nature of these 
objects and their relation to their more luminous counterparts.  

The intrinsic weakness of low-luminosity nuclei poses practical challenges on 
obtaining the data.  Aside from the issue of sensitivity,
often the main limitation stems from insufficient angular resolution
necessary to separate the faint central source from the galaxy background.
At virtually all wavelengths of interest the core emission constitutes only 
a small fraction of the total light, and hence contamination from the host
galaxy is severe.  To date, only a handful of objects have been adequately 
studied with multiwavelength observations, and even for these the 
wavelength coverage is sometimes highly incomplete and the data only 
approximate (NGC 1316 and NGC 3998: Fabbiano, Fassnacht, \& Trinchieri 1994; 
Sgr A$^*$: Narayan, Yi, \& Mahadevan 1995; M81: Ho, Filippenko, \& Sargent 
1996; NGC 4258: Lasota \etal 1996; M87: Reynolds \etal 1996; NGC 4594: 
Fabbiano \& Juda 1998, Nicholson \etal 1998).  Nonetheless, these studies 
already suggest that the SEDs of low-luminosity AGNs look markedly different 
compared to the SEDs normally seen in luminous AGNs.  The spectral 
peculiarities seen in low-luminosity AGNs hint at possibly significant 
departures in these objects from the standard AGN accretion disk model.

This paper presents SEDs for a small sample of low-luminosity AGNs for 
which reasonably secure black hole masses have been determined by 
dynamical measurements.  Data are presented for a total of seven 
objects (see Table 1 for a summary), four for the first time (NGC 4261, 
NGC 4579, NGC 6251, and M84); the SEDs for the remaining three (M81, M87, and 
NGC 4594) have been substantially updated and improved compared to previous 
publications.  The sample consists of four objects spectroscopically 
classified as low-ionization nuclear emission-line regions or LINERs 
(Heckman 1980; see Ho 1999a for a review), one Seyfert, and two objects 
that border on the definition of LINERs and Seyferts.  It is worth remarking 
that, although the sample is admittedly small and heterogeneous, it contains
{\it every} known low-luminosity object that has both a black hole mass 
determination and sufficient multiwavelength data to establish the 
SED.  The only two not included, namely Sgr A$^*$ and NGC 4258, have already 
been extensively discussed in the literature.  Because of the many complexities 
(\S\ 2) involved in defining the nuclear SEDs of these weak sources, because 
so few data of this kind exist in the literature, and because a companion 
paper (Ho \etal 1999b) relies critically on the details presented here, I devote 
considerable attention in \S\ 2 to describing the data selection for each 
object.   Section 3 summarizes the most noteworthy trends observed 
collectively in the sample, and section 4 discusses possible complications 
introduced by dust extinction; readers not interested in the details of each 
object may wish to skip directly to sections 3 and 4 for an overview of the 
main results.  Ho et al. (1999b) will present detailed modeling of these data 
based on accretion-disk calculations.

\section{Compilation of the Data}

As mentioned in \S\ 1, measuring the weak signal from low-luminosity AGNs 
requires observations that minimize the contamination from the bright 
background of the host galaxy.  High angular resolution, therefore, is 
indispensable at all wavelengths.  This study takes the following strategy 
for data selection.  (1) At radio wavelengths, only interferometric data will 
be used, preferably measured through VLBI techniques.  The radio jet 
component, if present, potentially can contaminate the core emission even in 
sub-arcsecond resolution.  (2) Data in the infrared (IR) window are at the 
moment most poorly constrained.  There are no useful far-IR data because nearly 
all the existing measurements have been obtained using the {\it Infrared 
Astronomical Satellite}, which has a beam \gax 1\amin.  Ground-based 
measurements in the mid-IR (10--20 \micron) and near-IR (1--3 \micron) 
are widely available in the literature, but these data should be regarded 
strictly as upper limits because of the relatively large apertures employed 
($\sim$3\asec--10\asec).  The energy distribution of normal stellar populations 
implies that the contamination from starlight in the near-IR should greatly 
exceed that in the mid-IR; thus, the mid-IR points should be more 
representative of the direct emission from the nucleus, although emission 
from hot ($\sim$100 K) dust grains can substantially boost the luminosity in 
this band (e.g., Willner \etal 1985).  (3) All of the optical and ultraviolet 
(UV) data are derived from observations made with the {\it Hubble Space 
Telescope (HST)}.  Photometry points generally pertain to an aperture of 
$\sim$0\farcs1, while apertures \lax 1\asec\ have been used for 
spectroscopic measurements.  (4) Finally, I quote only X-ray fluxes that 
originate from a nuclear source known to be compact at soft X-ray 
energies (0.5--2.5 keV).   The criterion for compactness, unless otherwise 
noted, is currently limited to the resolution of the High-Resolution Imager 
(HRI), approximately $\sim$5\asec, on either the {\it Einstein} or the 
{\it ROSAT} satellite.  Several objects have hard X-ray (2--10 keV) spectra 
acquired with {\it ASCA}, which has an angular resolution of $\sim$5\amin; 
HRI images show that the soft X-ray emission is compact in all these cases.

The following subsections describe the data chosen for each object.
Additional details can be found in Tables 2--8, and the individual 
SEDs are shown in Figures 1--7.  

\subsection{NGC 3031 (M81)}


The SED presented in Ho \etal (1996) has been updated (Table 2; Fig. 1) with 
new optical and UV photometry points from {\it HST}, an additional 
high-frequency radio point, and a more up-to-date hard X-ray spectrum  from 
{\it ASCA}.  The radio continuum between $\sim$1 and 15 GHz shows an inverted 
spectrum ($\alpha\,\approx$ --0.3 to --0.6, where $F_{\nu}\,\propto\,
\nu^{-\alpha}$).  It is difficult to assess the reality of the apparent slight 
turnover between 5 and 15 GHz because the two data points were not observed 
simultaneously; the radio core of M81 is highly variable on many timescales 
(Ho \etal 1999a).  The spectroscopically measured optical--UV slope of the 
featureless continuum is surprisingly steep.  Even after accounting for an 
estimated reddening of $E(B-V)$ = 0.094 mag, the slope is still $\sim$2 (Ho 
\etal 1996).  Devereux, Ford, \& Jacoby (1997) recently obtained an {\it HST} 
WFPC2 image of the nucleus of M81 using a filter centered near 1500 \AA.  They 
obtained a flux $\sim$4 times higher than that reported in the Faint Object 
Spectrograph (FOS) UV spectrum of Ho \etal (1996).  The two new points based 
on optical images (Bower \etal 1996; Devereux \etal 1997), on the other hand, 
are consistent with the spectroscopic measurements.  Maoz \etal (1998) have 
reanalyzed the FOS UV data and concluded that the nucleus may have been 
miscentered in the aperture during the observations.  This may account for the 
discrepancy between the FOS and WFPC2 fluxes, although it cannot be excluded 
that the nucleus varied in the UV between the two observing epochs.  Adopting 
the WFPC2 UV flux, the optical-UV slope is now $\alpha\,\approx$ 1.3--1.4.  
The nucleus of M81 emits a nonthermal continuum in the hard X-ray band.  The 
spectrum between 2--10 keV is well described by a single power law with a 
slope of $\alpha$ = 0.85$\pm$0.04 (Ishisaki \etal 1996; see also Serlemitsos, 
Ptak, \& Yaqoob 1996), similar to that seen in luminous Seyfert 1 nuclei 
(Turner \& Pounds 1989; Nandra \etal 1997); the luminosity in this case, 
however, is just 2\e{40} \lum.  
It is noteworthy that M81 emits substantially 
more energy in the X-rays relative to the UV than in luminous AGNs.  The 
two-point spectral index between 2500 \AA\ and 2 keV, $\alpha_{\rm ox}$, is 
1.08, smaller than the average value in quasars (1.4) or in luminous Seyfert 
1 nuclei (1.2) (Mushotzky \& Wandel 1989).

\subsection{NGC 4261 (3C 270)}


Figure 2 (Table 3) displays the SED of the nucleus.  The radio core is 
resolved even on a scale of several milliarcseconds.  No obvious point source 
can be seen in the 1.6 and 8.4~GHz maps of Jones \& Wehrle (1997), so I take 
as a reasonable approximation the peak intensities at these frequencies.  The 
resulting radio spectrum has $\alpha$ = 0.   The three optical points derived 
from WFPC2 images (Ferrarese, Ford, \& Jaffe 1996) 
define a slope of $\alpha\,\approx$ 2.2.  The FOS spectrum of Ferrarese et 
al., taken with a similar-sized aperture (0\farcs1), suggests that most of the 
continuum is nonstellar.  The very steep optical continuum probably results 
mainly from reddening by dust internal to NGC 4261 and most likely associated 
with the nuclear disk.  Patchy extinction can be seen throughout the
disk and in the immediate vicinity of the nucleus (Ferrarese et al. 1996).
Dust is also the most likely culprit for the complete absence of UV emission:
the upper limit to the flux at 2300 \AA\ is 10$^{-16}$ \flux\ (Zirbel \& Baum
1998), a factor of 80 lower than expected from a simple power-law
extrapolation of the observed optical continuum.  Estimating the intrinsic 
optical and UV luminosity of the nucleus, however, is difficult without 
prior knowledge of the absorbing column, the source geometry, and the 
extinction law (see \S\ 4).  The extinction due to the Galaxy is small.  The 
foreground hydrogen column density is only $N_{\rm H}$ = 1.6\e{20} \percm2,
which, for the conversions $E(B-V)\,=\,N_{\rm H}$/(5.8\e{21} cm$^{-2}$) mag 
and $A_V/E(B-V)$ = 3.1 (Bohlin, Savage, \& Drake 1978), translates to 
$A_V$ = 0.084 mag (Table 1).   A rough estimate of the magnitude of the 
internal extinction can be obtained from the Balmer decrement observed through 
a larger aperture (2\asec$\times$4\asec) by Ho \etal (1997b), H\al/H\bet\ = 
4.9.\footnote{Ferrarese \etal (1996) quote a significantly larger value of 
H\al/H\bet\ = 9.7.  Comparison of their relative line 
intensities with those of Ho \etal (1997b) suggests that they have 
underestimated the intensity of H\bet\ by a factor of $\sim$2.  Ferrarese 
et al. could not remove the underlying stellar absorption lines from 
their spectra because of the limited signal-to-noise ratio of their data.  
This effect preferentially biases the Balmer line intensities, H\bet\ more so 
than H\al, to low values.}  After removing the Galactic contribution using the 
extinction law of Cardelli, Clayton, \& Mathis (1989), H\al/H\bet(internal) = 
4.8; assuming the Galactic extinction law and a Case B$^{\prime}$ intrinsic 
H\al/H\bet\ of 3.1, which is thought to be appropriate for the conditions 
in the narrow-line regions of AGNs (e.g., Gaskell \& Ferland 1984), 
$A_V$(internal) = 1.4 mag.  This value is much higher than that inferred 
from the soft X-ray observations of Worral \& Birkinshaw (1994), who placed a  
limit of $N_{\rm H}\,<\,$ 3.9\e{20} \percm2, or $A_V\,<$ 0.2 mag for a normal 
dust-to-gas ratio.  

Worral \& Birkinshaw (1994) observed NGC 4261 with the {\it ROSAT} 
Position-Sensitive Proportional Counter (PSPC).  Despite the coarse angular 
resolution of the PSPC ($\sim$25\asec), they found that $\sim$50\% of the flux 
comes from a spatially unresolved component whose spectrum is well fitted 
by a power-law function.  The nonthermal component emits $L$(0.2--1.9 keV) 
= 4.7\e{40} \lum.  Inspection of an archival {\it ROSAT} HRI 
image confirms that most of the soft X-ray emission indeed does stem from a 
compact component.   The uncertain impact of dust extinction on the 
optical-UV continuum renders estimates of $\alpha_{\rm ox}$ highly 
unreliable.  If $A_V$(internal) = 0 mag (correcting only for 
the Galactic contribution), the optical slope is $\alpha$ = 2.06, which, 
when extrapolated to 2500 \AA\, yields $\alpha_{\rm ox}$ = 0.44. 
Choosing $A_V$(internal) = 1.4 mag and the Galactic extinction law, the 
optical slope becomes $\alpha$ = 0.67, and $\alpha_{\rm ox}$ = 0.84.  These 
two extremes probably bracket the true value.

\subsection{NGC 4374 (M84, 3C 272.1)}


Jones, Terzian, \& Sramek (1981) mapped the nucleus with a resolution of 4 mas 
at 1.6~GHz.  The morphology of the radio core can be modeled by an elliptical 
Gaussian with dimensions 1.4 mas $\times$ 6.0 mas, presumably because a 
parsec-scale jet still contributes to the emission on these scales.  The radio 
flux from the accretion flow, therefore, ought to be less than indicated in 
Figure 3 (Table 4).   At optical wavelengths the nucleus appears unresolved in 
WFPC2 images.   Bower \etal (1997) measure $V$ = 19.9 mag and $(V-I)$ = 
1.6 mag.   The stellar contribution to the pointlike nucleus is unclear, but 
judging from the complete absence of stellar absorption features between 6300 
\AA\ and 6800 \AA\ in spectra acquired through a 0\farcs2 slit (Bower \etal 
1998), most of the light is probably nonstellar.   The ground-based spectrum 
in Ho \etal (1995), for instance, shows a very noticeable (equivalent width 
$\sim$1 \AA) absorption line at \lamb\ = 6495 \AA\ due to 
\ion{Ca}{1}+\ion{Fe}{1}.  The $(V-I)$ color of the nucleus, which corresponds 
to $\alpha$ = 3.5, again suggests substantial reddening by dust, as does the 
nondetection of UV emission (Zirbel \& Baum 1998); the images in Bower \etal 
(1997), in fact, clearly show dust patches projected on the front of the 
nucleus.  These authors used the $(V-I)$ color map and an assumed intrinsic 
$(V-I)$ color for ellipticals to estimate a mean internal extinction of $A_V$ 
= 0.54 mag within a relatively large region of 14\farcs5$\times$8\farcs4.  
Close to the nucleus the extinction could be higher than this value.  The 
intrinsic Balmer decrement measured through a 2\asec$\times$4\asec\ aperture 
around the nucleus, for instance, is H\al/H\bet\ = 4.7 (Ho \etal 1997b), or 
$A_V$ = 1.3 mag.  The optical and UV points were dereddened using this $A_V$.  

The {\it Einstein} HRI image of M84 reveals a relatively complex central 
morphology with significant extended emission (Fabbiano, Kim, \& Trinchieri 
1992); the extended structure is also clearly evident in an archival 
{\it ROSAT} HRI image.  The 0.5--4 keV luminosity given by Fabbiano \etal 
(1992), 5.3\e{40} \lum\ (assuming a ``Raymond-Smith'' thermal spectrum with 
$kT$ = 1 keV and a line-of-sight $N_{\rm H}$ = 1.7\e{20} \percm2), should be 
taken only as an upper limit to the luminosity of a pointlike nucleus.  
Extrapolating the $\alpha\,\approx$ 3.1 optical slope to the UV gives an 
estimate of $\alpha_{\rm ox}\,\approx$ 0.75.

\subsection{NGC 4486 (M87, 3C 274)}


Reynolds \etal (1996) presented a sparse SED for the nucleus.  Figure 4 (Table 
5) gives an updated version that includes additional data from the IR to UV 
region.  As in the objects discusses so far, the radio spectrum is either flat 
or slightly inverted; the VLBI points (filled symbols) have a mean spectral 
index $<\alpha>\,\approx\,0.1$.  The apparent change in the spectral index 
from one point to another reflects the nonuniform resolution of the four 
experiments, which results in varying degrees of contamination from the 
jet, the nonsimultaneity of the observations, or both.  It is instructive to 
note that the jet in this case introduces very substantial contamination to 
the radio core emission even on a scale of $\sim$1\asec.  The open symbols 
show the three VLA points from Biretta, Stern, \& Harris (1991); clearly they 
lie systematically higher by about a factor of 10 compared to the VLBI flux 
densities.  The optical (6800 \AA) to UV (1200 \AA) spectrum, observed 
nearly simultaneously with the {\it HST}/FOS through the same (0\farcs2) 
aperture, traces a smooth, continuous curve that can be well fitted with a 
double power law ($\alpha$ = 1.75 for \lamb \lax 4500 \AA, $\alpha$ = 1.41 
for \lamb \gax 4500 \AA; Tsvetanov \etal 1998).  The two $I$-band points 
derived from archival {\it HST} images agree very well with the FOS spectrum 
(within $\pm$10\%).  On the other hand, the point taken from the Faint 
Object Camera (FOC) measurement of Maoz \etal (1995) is about 50\% higher 
than predicted from the FOS spectrum; variability is a plausible explanation 
for this discrepancy.  The fluxes in the $J$ (1.25\micron) and $K$ (2.2\micron)
bands (Stiavelli, Peletier, \& Carollo 1997), both acquired under sub-arcsecond 
seeing conditions, also appear to follow the extrapolated optical power law. 
Taken at face value, the 10\micron\ point deviates quite strongly from the 
power law at shorter wavelengths, but the relatively large aperture of the 
observations (6\asec) allows room for substantial contamination.  The soft 
X-ray flux recorded by the {\it Einstein} HRI is $\sim$50\% lower than that 
seen in the {\it ROSAT} HRI [$L$(1 keV) = 5.4\e{40} \lum, assuming a power-law 
spectrum with $\alpha$ = 0.7 and a line-of-sight $N_{\rm H}$ = 2.5\e{20} 
\percm2; see Reynolds \etal 1996].  This level of long term X-ray variability 
is not uncommon in low-luminosity AGNs (Serlemitsos \etal 1996).  An 
{\it ASCA} spectrum of M87 is available, but the nucleus could not be clearly 
detected in the relatively short exposure (Reynolds \etal 1996).  Combining 
the UV spectrum with the average of the two HRI fluxes, $\alpha_{\rm ox}$ = 
1.06.

M87 belongs to the minority of LINERs ($\sim$25\%; Maoz \etal 1995; Barth 
\etal 1998) that shows prominent UV emission.  The optical-UV continuum of M87 
evidently suffers little or no internal extinction.  The best-fitting 
double power-law model shown in Figure 4 requires only $A_V$ = 0.12 mag 
for a Milky Way extinction law (Tsvetanov \etal 1998), very similar to the
Galactic contribution of $A_V$ = 0.078 mag.  Note that the Balmer decrement 
measured by Ho \etal (1997b) in a 2\asec$\times$4\asec\ aperture yields a
substantially larger $A_V$ of 1.0 mag.  This example underscores the 
dangers of comparing observations taken with markedly different apertures. 

\subsection{NGC 4579 (M58)}


The 2.3-GHz and 8.4-GHz interferometric observations of Sadler \etal (1995), 
both made with a beam of 0\farcs03, define a rising radio spectrum with 
$\alpha$ = --0.19 (Fig. 5; Table 6).  The {\it HST}/FOS UV spectrum of Barth 
\etal (1996) extends from $\sim$3300 \AA\ to 1150 \AA; excluding a broad 
feature near 3000 \AA\ due to Balmer continuum and \feii\ emission, the 
spectrum can be described by a nearly featureless power-law function, 
$F_{\nu}\,\propto\,\nu^{-1}$.  There is tentative evidence that the nucleus 
varies in the UV.  The 2300 \AA\ flux in the FOS spectrum is a factor of 3 
lower than that reported by Maoz \etal (1995) based on an FOC image. In either 
case, the UV output, again, is quite low with respect to the X-rays: 
$\alpha_{\rm ox}$ = 0.78 and 1.02 for the FOS and FOC UV flux, respectively.
A compact, nonthermal component dominates the {\it ASCA} spectrum of 
NGC 4579.  Terashima \etal (1998) find that the 2--10 keV continuum can 
be modeled as a moderately absorbed (intrinsic $N_{\rm H}$ = 4\e{20} \percm2)
power law with $\alpha$  = 0.72 and luminosity 1.5\e{41} \lum.  
The value of $N_{\rm H}$ derived from the X-ray spectrum roughly matches the 
intrinsic extinction estimated by Barth \etal (1996) based on the shape of the 
observed UV continuum.

\subsection{NGC 4594 (M104, Sombrero)}


Fabbiano \& Juda (1997) presented an approximate SED for the nucleus of 
NGC 4594.  The optical and UV points used in that study, however, were highly 
uncertain.  The $B$-band data, based on the pre-refurbishment FOC image of 
Crane \etal (1993), may have been underestimated because of the nonlinear 
behavior of the FOC (see Crane \& Vernet 1997).  On the other hand, the UV 
flux taken with the very large aperture (10\asec$\times$20\asec) of the {\it 
International Ultraviolet Explorer} is undoubtedly far too high.  Here the SED 
of NGC 4594 is reassessed, paying close attention to aperture effects in the 
difficult optical-UV region; a preliminary version of these data appears in 
Nicholson \etal (1998).
 
The high-resolution data between 0.6 and 15 GHz define a compact flat-spectrum 
core with $\alpha$ ranging from 0.2 to --0.4 (Fig. 6; Table 7).  The 
nonstellar component of the optical and UV continuum, on the other hand, is 
somewhat difficult to specify.  The optical FOS spectrum of Kormendy \etal 
(1997), taken through a 0\farcs21 aperture, has a very red continuum 
($\alpha\,\approx\,3.5$).  The steep optical slope is most likely caused by an 
increasing contamination of starlight toward longer wavelengths (even within 
such a small aperture) rather than by dust reddening of a purely featureless 
continuum.  The dilution of the depth of the stellar absorption lines 
indicated to Kormendy et al. that approximately 50\% of the light at $B$ 
within their 0\farcs21 aperture comes from the nonstellar continuum.   This 
roughly agrees with the strength of the point source extracted from an 
archival WFPC2 $V$-band image (Table 7).  The point-source luminosity 
similarly measured from an $I$-band image, however, falls significantly 
below (by a factor $\sim$4) the extrapolated FOS spectrum if one assumes 
the degree of stellar contamination to be constant between $B$ and $I$.  
This suggests that the nonstellar component contributes much less to the 
red end of the FOS spectrum ($\sim$6800 \AA), probably on the order of 
25\% or so.   The $I$-band flux is therefore adopted for the red end of the 
FOS spectrum.  If one considers the two photometry points to be a reliable 
measure of the nonstellar component, the optical slope decreases to 
$\alpha$ = 1.3.  The UV spectrum, taken with an even larger aperture of 
0\farcs86, likewise suffers from wavelength-dependent starlight contamination 
(Nicholson \etal 1998).  I excluded from the SED the portion of the spectrum 
longward of \lamb = 2200 \AA, where incipient stellar absorption features and 
a sharply rising continuum suggest a sizable contribution from stars.  The 
final adopted optical and UV points (shown as filled symbols) are well joined 
by a power law with $\alpha\,\approx$ 1.5.

The {\it ASCA} 2--10 keV spectrum is predominantly nonthermal.  Nicholson 
\etal (1998) obtained a best fit with a power law with $\alpha$ = 0.63 
and $L$(2--10 keV) = 1.1\e{41} \lum.
The morphology of the central region 
as seen in the {\it ROSAT} HRI image (Fabbiano \& Juda 1997) indicates that 
most of the hard X-ray emission is likely to originate from the nucleus.  
The X-ray band energetically dominates over the UV band, with $\alpha_{\rm ox}$ 
= 0.89.  The average extinction inferred from the Balmer decrement on arcsecond 
scales appears modest ($A_V$ = 0.25 mag; Ho \etal 1997b) and roughly agrees 
with the intrinsic hydrogen column derived from the {\it ASCA} spectrum 
($N_{\rm H}$ = 5.3\e{20} \percm2\ or $A_V$ = 0.28 mag).

\subsection{NGC 6251}


Good quality VLBI maps are available to isolate the core radio emission 
(Cohen \& Readhead 1979; Jones \etal 1986), but the nonsimultaneous nature of 
the observations makes definition of the intrinsic radio spectrum ambiguous 
(Fig. 7; Table 8). The two low-frequency points, for instance, yield a much 
steeper spectral index than the two high-frequency points ($\alpha$ = 0.4 
compared to $\alpha$ = --1.2).  The only conclusion that can be drawn at 
this stage is that the radio spectrum, as in all the sources studied here, is 
consistent with that of a self-absorbed synchrotron source.   Four {\it HST} 
photometry points have been measured for the pointlike nucleus, corresponding 
roughly to the $U, B, V$, and $I$ bands (Crane \& Vernet 1997).   Taking into 
consideration the nonlinear effects of the FOC that affected the $U$ and $B$ 
fluxes, the optical to near-UV slope is consistent with $\alpha\,\approx\, 
1.7$.  Combining the extrapolated UV flux with the soft X-ray (PSPC) flux 
reported by Worral \& Birkinshaw (1994), $\alpha_{\rm ox}$ = 0.83.  
Birkinshaw \& Worral (1993) performed a detailed analysis of the  PSPC data 
and concluded that nearly all ($\sim$90\%) of the emission arises from 
a spatially unresolved, power-law component with a diameter \lax 4\asec.  More 
detailed spectral information comes from Turner et al.'s (1997)  statistical 
study of the {\it ASCA} spectra of a large sample of Seyfert 2 nuclei, which 
included NGC 6251.   The best-fitting model found by Turner et al. requires 
a thermal (Raymond-Smith) plasma with a temperature of $kT$ = 0.85 keV 
added to a power-law component characterized by $\alpha$ = 
1.11$^{+0.16}_{-0.19}$, $L$(2--10 keV) = 1.3\e{42} \lum, and an intrinsic 
$N_{\rm H}$ = 1.4\e{21} \percm2.  
The absorbing column obtained from the X-rays predicts a 
sizable internal extinction of $A_V$ = 0.75 mag.  The Balmer decrement given 
in Shuder \& Osterbrock (1981) requires a much larger $A_V\,\approx$ 5 mag, 
although the reported H\bet\ intensity appears to be rather uncertain. 

\section{General Properties of the SEDs}

Luminous AGNs generally display a fairly ``universal'' SED (see, e.g., Sanders 
\etal 1989 and Elvis \etal 1994).  The continuum from the IR to the 
X-rays, roughly flat in log~$\nu F_{\nu}$--log~$\nu$ space, can be represented 
by an underlying power law with $\alpha\,\approx$ 1 superposed with several 
distinct components, the most prominent of which is a broad UV excess.  This 
so-called big blue bump is conventionally interpreted as thermal emission 
arising from an optically thick, geometrically thin accretion disk (Shields 
1978; Malkan \& Sargent 1982).  The largest spectral difference among AGNs 
manifests itself in their brightness in the radio band --- a factor of nearly 
$10^2$--$10^3$ in radio power distinguishes  ``radio-loud'' from 
``radio-quiet'' objects.  

The broad-band spectra of the seven low-luminosity AGNs presented here share a 
number of common traits, and yet they differ markedly from the SEDs of 
luminous AGNs.  To illustrate this point, Figure 8 compares the SEDs of the 
present sample with the median SED of radio-loud and radio-quiet luminous AGNs 
taken from Elvis \etal (1994); all the curves have been normalized at 1 keV.  
Several features of the low-luminosity SEDs are noteworthy:  

(1) The optical-UV slope is quite steep.  The power-law indices for the seven 
objects average $\alpha\,\approx\,1.8$ (range 1.0--3.1; see Table 9), or 1.5 
if the possibly highly reddened objects NGC 4261 and M84 are excluded, whereas 
in luminous AGNs $\alpha \,\approx$ 0.5--1.0.  

(2) The UV band is exceptionally dim relative to the optical and X-ray bands.
There is no evidence for a big blue bump component in any of the objects.
Indeed, the SED reaches a local minimum somewhere in the far-UV or extreme-UV 
region.  This leads to the above-mentioned steep optical-UV slope and to 
systematically low values of $\alpha_{\rm ox}$.  Table 9 gives 
$<\alpha_{\rm ox}>\,\approx\,0.9$, to be compared with $<\alpha_{\rm ox}>$ = 
1.2--1.4 for luminous Seyferts and QSOs (Mushotzky \& Wandel 1989).  In other 
words, the low-luminosity AGNs in the present sample, most of which are LINERs, 
are systematically ``X-ray loud'' (relative to the UV) compared to AGNs 
of higher luminosity.  This modification of the SED from UV to X-ray energies
leads to a harder ionizing spectrum, and it offers an explanation, 
at least in part, for the characteristically lower ionization state of the 
emission-line regions (Ho 1999b). 

(3) The hard X-ray (2--10 keV) spectra, where available, are well fitted with 
a power-law function with $\alpha\,\approx$ 0.6--0.8, very similar to spectra 
observed in high-luminosity sources. 


(4) There is tentative evidence for a maximum in the SED at mid-IR or longer 
wavelengths.  Despite the relatively large apertures employed in the mid-IR 
observations, the 10-\micron\ point should be largely uncontaminated by 
starlight, although dust emission could contribute significantly in this 
band. 

(5) The nuclei have radio spectra that are either flat or inverted.  The 
radio brightness temperature, where available, reach at least 
10$^{9}$--10$^{10}$ K.  The radio cores, therefore, are self-absorbed 
synchrotron sources.

(6) One usually gauges the degree of radio dominance in AGNs by the ratio 
of the specific luminosities in the radio to the optical band.  For instance,
Kellermann \etal (1989) classify the radio strength of QSOs by the parameter 
$R\,\equiv\,F_{\nu}({\rm 6 cm})/F_{\nu}(B)$; radio-quiet members have 
$R\,\approx$ 0.1--1, and radio-loud members are distinguished by $R$ \gax 100.
Adopting the same criterion, {\it all} of the objects, including the three 
spiral galaxies in the sample (M81, NGC 4579, and NGC 4594), qualify as being 
{\it radio-loud}.  M81 has the smallest radio-to-optical ratio, but $R$ is 
still $\sim$50.  This finding runs counter to the usual notion that only 
elliptical galaxies host radio-loud AGNs.  Note that if the total (host galaxy 
+ nucleus) $B$ luminosity had been used, which in these sources significantly 
exceeds the nuclear value alone, all the objects, with the possible exception 
of NGC 6251, would have been considered radio quiet.  

(7) The sample of objects studied here is intrinsically extremely faint.  
To cast this statement in a more familiar context and to fully appreciate the 
enormous challenge in detecting these objects, we note that AGNs that 
occupy the upper end of the luminosity distribution, namely QSOs, 
typically have nonstellar continua with absolute magnitudes --30 $<$ 
$M_B^{nuc}$ $<$ --23.  Classical Seyfert nuclei, such as those from the 
Markarian survey, are characterized by --23 $<$ $M_B^{nuc}$ $<$ --18.   By 
contrast, the nonstellar nuclear magnitudes listed in Table 9 lie in the range 
--14.7 $<$ $M_B$ $<$ --8.9.  Excluding the two extreme cases, 
$< M_B^{nuc} >$ = --11.5 mag.  The host galaxies themselves, on the other 
hand, are luminous $L^*$ systems ($< M_{B_T}^{0} >\,\approx$ --21.1 mag; 
Table 1), and hence the nuclei comprise merely $\sim$0.01\% of the 
total optical light of the host galaxies.

(8) The bolometric luminosities of the sources (Table 9), obtained by 
integrating the power-law segments shown on Figure 8, range from 
$L_{\rm bol}$  = 2\e{41} to 8\e{42} \lum, or $\sim 10^{-6}-10^{-3}$ times the 
Eddington rate for the black hole masses listed in Table 1.  The bolometric 
luminosities would be lower if the mid-IR peak has been overestimated, but 
this would not affect the conclusion that the Eddington ratios are very low.
The X-ray band, arbitrarily defined here as the region 
from 0.5 to 10 keV, carries 6\%--33\% of $L_{\rm bol}$.  

\section{Uncertainties due to Dust Extinction}

Important aspects of the interpretation of the data depend on the 
intrinsic luminosity of the UV region, as this impacts conclusions concerning 
the optical-UV continuum slope, the presence of the big blue bump, and the
strength and shape of the ionizing spectrum.  The UV bandpass, unfortunately,
is strongly affected by dust extinction.  Although the effects of
extinction can, in principle, be corrected, in practice such a procedure
is fraught with a number of uncertainties, which I briefly mention here.

First, it is unclear how to measure the amount of dust affecting the
UV continuum source.  Two measures of extinction are traditionally used.
One is based on comparison of the observed ratio of a pair of emission lines
with their intrinsic ratio, after adopting a form for the extinction law.
The intrinsic spectrum of the hydrogen recombination lines is well known for
conditions prevailing in the low-density narrow-line regions of AGNs, and
the derived optical extinction does not depend sensitively on the exact form
of the extinction law.  A second method uses the neutral hydrogen column
density derived from X-ray spectra to calculate the absorbing
column under the assumption that the dust-to-gas ratio and the grain
properties are the same as those in the Galaxy.  The latter assumptions,
however, need not hold, especially in the vicinity of an AGN where grains
can be destroyed by the harsh radiation field (Voit 1991).  Low values of
$N_{\rm H}$ derived from X-ray observations also do not necessarily imply low
extinction because some dust can be associated with the ionized medium.  In 
either case, a major source of uncertainty is whether the UV continuum
source traverses through the same absorbing medium as probed by the X-rays 
or by the optical emission lines.  The extinction obtained from the X-ray 
absorbing column in AGN spectra, for instance, often greatly exceeds the 
extinction inferred from the Balmer decrement (e.g., Reichert \etal 1985).

One can attempt to estimate directly the extinction of the continuum by
searching for spectral signatures imprinted by the assumed extinction law.
For a Galactic extinction law, the most noticeable feature in the mid-UV is a
broad depression centered near 2200 \AA.   The UV spectra of AGNs,
on the other hand, usually do not show this feature (McKee \& Petrosian
1974; Neugebauer \etal 1980; Malkan \& Oke 1983; Tripp, Bechtold, \& Green 
1994) despite independent evidence
for dust from other indicators (e.g., emission-line ratios).  This result
implies one or more of the following possibilities: (1) the UV continuum and
the line-emitting gas experience different amounts of extinction because
the distribution of dust along the line-of-sight is patchy; (2) the strong
radiation field destroys dust grains close to the central continuum source;
and (3) the extinction law in extragalactic environments differs from that of 
the Galaxy, specifically in having a much weaker 2200-\AA\ bump.  Support for 
the latter possibility comes from observations of starburst galaxies (Calzetti, 
Kinney, \& Storchi-Bergmann 1994) and of the Small Magellanic Cloud (SMC; 
Bouchet \etal 1985) whose UV spectra generally show a very weak, if any, 
2200-\AA\ bump.  Gordon, Calzetti, \& Witt (1997) argue that the absence of the 
2200-\AA\ feature in these systems is inherent in their extinction law and not 
merely a consequence of geometry effects.  The properties of dust grains, 
specifically the small graphite grains responsible for the 2200-\AA\ bump 
(Mathis 1994), evidently can be greatly affected by environmental conditions 
such as star-formation activity and/or metallicity.  It is not difficult to 
imagine that the same may be the case in the vicinity of an AGN, where the 
extinction law can be modified dramatically by the intense radiation field 
(Laor \& Draine 1993; Czerny \etal 1995).

Is the apparent faintness of the UV band intrinsic to the SEDs or is it 
instead simply a consequence of dust extinction?  Luminous AGNs have an 
optical-UV slope of $\alpha_{\rm ou}\,\approx$ 0.5--1.0, whereas in the 
present sample $\alpha_{\rm ou}\,\approx$ 1.5--2.0.  Let us estimate how 
much UV extinction is required to redden the ``typical'' AGN spectrum to that 
seen here.  Following Maoz \etal (1998), I consider the extinction curve of 
the SMC (Bouchet \etal 1985), as parameterized by Pei (1992), and the empirical 
starburst ``attenuation curve'' of Calzetti \etal (1994), as parameterized by 
Calzetti (1997).  Both curves lack a 2200-\AA\ bump, but the Calzetti et al. 
curve is greyer.  Adopting a fiducial optical and UV wavelength of 6500 \AA\ 
and 2500 \AA, respectively, reddening $\alpha_{\rm ou}$ by a slope of 1 
requires $A_V$ = 0.6 mag and $A_{2500}$ = 1.5 mag for the SMC curve.  The 
shallower Calzetti et al. curve gives $A_V$ = 1.1 mag and $A_{2500}$ = 2.0 
mag.  For a more extreme case of $\Delta \alpha_{\rm ou}$ = 2, $A_V$ = 1.3 
mag and $A_{2500}$ = 3.0 mag for the SMC curve, and $A_V$ = 2.3 mag and 
$A_{2500}$ = 4.1 mag for the Calzetti et al. curve.  These cases are 
illustrated in Figure 9.  Given the limited accuracy of the available data and 
the unknown form of the extinction law in AGNs, it is difficult to rule out 
dust reddening as the principal cause of the observed weakness of the 
UV band, although one can probably exclude Galactic-type dust grains 
from the absence of the predicted strong 2200-\AA\ feature (thin solid lines 
in Fig. 9).  In fact, the necessary amount of visual extinction, 
$A_V\,\approx$ 1--2 mag, is not unreasonable compared to those obtained from 
the Balmer decrements (Table 1).  Note that an average correction of 
$A_{2500}$ = 2.0 mag will change $\alpha_{\rm ox}$ from 0.9 to 1.2, as seen 
in luminous Seyfert 1s.  

Nonetheless, in order to account for the {\it systematic} differences between 
the SEDs of the two luminosity classes, one would have to postulate that 
low-luminosity AGNs {\it systematically} have either greater levels of 
extinction or a different dust extinction law compared to high-luminosity 
AGNs.  The first explanation is not supported by the data, at least insofar as 
Balmer decrements can be used to gauge the continuum extinction; the 
internal extinctions listed in Table 1 do not appear anomalous compared to 
those found in more luminous Seyfert galaxies (e.g., Shuder \& Osterbrock 
1981; Dahari \& De Robertis 1988).  The second hypothesis is ad hoc and 
difficult to test experimentally.  Moreover, if exposure to the AGN 
environment indeed does modify the extinction law, one would na\"\i vely expect 
the effect to be greatest on the most powerful AGNs, exactly the opposite 
of what is observed.  Thus, although dust extinction can in 
principle be responsible for the spectral peculiarities seen in low-luminosity 
objects, the alternative view that the nonstandard SEDs are intrinsic to the 
sources is also tenable and probably more favorable.

\section{Summary}

Broad-band spectra are presented for seven low-luminosity AGNs.  Although the 
sample is still limited and heterogeneous, this is the most extensive effort 
so far to systematically investigate the SEDs of these weak sources.  
Contamination by emission from the host galaxy can severely corrupt the faint 
signal from the nucleus, and careful attention has been paid to select only 
the highest quality nuclear fluxes in assembling the SEDs.  A comparative 
study of the gross features of the SEDs reveals that the SEDs of 
low-luminosity AGNs, as a group, look strikingly different compared to the 
standard energy spectrum of luminous Seyfert galaxies and QSOs.  Most of the 
differences stem from the exceptional faintness of the UV continuum in the 
low-luminosity objects.  The so-called big blue bump is very weak or 
altogether absent, thereby making the continuum shape between optical and UV 
wavelengths steeper than normal and the X-ray band energetically more 
important.  
Extinction by dust grains with properties different from those of Galactic 
composition may be responsible for suppressing the UV emission in 
low-luminosity AGNs, but an alternative, more intriguing possibility is that 
the absence of the big blue bump is a property intrinsic to these sources.  
Another notable property of the SEDs is the prominence of the 
compact, flat-spectrum radio component relative to the emission in other 
energy bands.  All seven nuclei in the sample, including the three hosted by 
spiral galaxies, can be considered ``radio-loud.''   The AGNs investigated 
here indeed are very quiescent objects.  Their bolometric luminosities range 
from 2\e{41} to 8\e{42} \lum, which correspond to an Eddington ratio of 
$\sim 10^{-6}-10^{-3}$.



\acknowledgments
L.~C.~H. acknowledges financial support from a Harvard-Smithsonian Center for
Astrophysics postdoctoral fellowship, from NASA grant NAG 5-3556, and from NASA
grants GO-06837.01-95A and AR-07527.02-96A from the Space Telescope
Science Institute (operated by AURA, Inc., under NASA contract NAS5-26555).
I thank Chien Peng and Edward Moran for assistance in analyzing some of the
{\it HST} and {\it ROSAT} HRI images, respectively, and I am grateful to
Marianne Vestergaard for valuable comments on an earlier version of the paper.
I thank Ramesh Narayan for discussions on the theoretical interpretation of the
data.  This work made extensive use of the NASA/IPAC Extragalactic Database
(NED) which is operated by the Jet Propulsion Laboratory, California Institute
of Technology, under contract with NASA.


\clearpage

\vskip 0.3truein
\centerline{\bf{References}}
\medskip

\refindent 
B\"{a}\"{a}th, L.~B., \etal 1992, \aa, 257, 31

\refindent 
Bartel, N., \etal 1982, \apj, 262, 556

\refindent 
Barth, A.~J., Ho, L.~C., Filippenko, A.~V., \& Sargent, W.~L.~W. 1998, \apj,
496, 133

\refindent 
Barth, A.~J., Reichert, G.~A., Filippenko, A.~V., Ho, L.~C.,
Shields, J.~C., Mushotzky, R.~F., \& Puchnarewicz, E.~M. 1996, \aj, 112, 1829

\refindent 
Bash, F.~N., \& Kaufman, M. 1986, \apj, 310, 621

\refindent 
Biretta, J., Stern, C.~P., \& Harris, D.~E. 1991, \aj, 101, 1632

\refindent 
Birkinshaw, M., \& Worrall, D.~M. 1993, \apj, 412, 568

\refindent
Blandford, R.~D., \& Rees, M.~J. 1992, in Testing the AGN Paradigm, ed.
S. Holt, S. Neff, \& M.  Urry (New York: AIP), 3

\refindent 
Bohlin, R.~C., Savage, B.~D., \& Drake, J.~K. 1978, \apj, 224, 132

\refindent 
Bouchet, P., Lequeux, J., Maurice, E., Prevot, L., \& Prevot-Burnichon,
M.~L. 1985, \aa, 149, 330

\refindent 
Bower, G.~A., \etal 1998, \apj, 492, L111

\refindent 
Bower, G.~A., Heckman, T.~M., Wilson, A.~S., \& Richstone, D.~O. 1997, \apj,
483, L33


\refindent 
Bower, G.~A., Wilson, A.~S., Heckman, T.~M., \& Richstone, D.~O. 1996, \aj,
111, 1901

\refindent 
Calzetti, D. 1997, in The Ultraviolet Universe at Low and High Redshift, ed.
W.~H. Waller et al. (New York: AIP), 403

\refindent 
Calzetti, D., Kinney, A.~L., \& Storchi-Bergmann, T. 1994, \apj, 429, 582

\refindent 
Cardelli, J.~A., Clayton, G.~C., \& Mathis, J.~S. 1989, \apj, 345, 245

\refindent 
Cohen, M.~H., \& Readhead, A.~C.~S. 1979, \apj, 233, L101

\refindent
Crane, P., \etal 1993, \aj, 106, 1371

\refindent 
Crane, P., \& Vernet, J. 1997, \apj, 486, L91

\refindent 
Czerby, B., Loska, Z., Szczerba, R., Cukierska, J., \& Madejski, G. 1995,
AcA, 45, 623

\refindent 
Dahari, O., \& De Robertis, M.~M. 1988, \apjs, 67, 249

\refindent 
de Vaucouleurs, G., de Vaucouleurs, A., Corwin, H.~G., Jr., Buta, R.~J.,
Paturel, G., \& Fouqu\'e, R. 1991, Third Reference Catalogue of Bright
Galaxies (New York: Springer)

\refindent 
Devereux, N.~A., Becklin, E.~E., \& Scoville, N. 1987, \apj, 312, 529

\refindent 
Devereux, N.~A., Ford, H.~C., \& Jacoby, G. 1997, \apj, 481, L71

\refindent 
Elvis, M., \etal 1994, \apjs, 95, 1

\refindent 
Fabbiano, G., Fassnacht, C., \& Trinchieri, G. 1994, \apj, 434, 67

\refindent 
Fabbiano, G., \& Juda, J.~Z. 1997, \apj, 476, 666

\refindent 
Fabbiano, G., Kim, D.-W., \& Trinchieri, G. 1992, \apjs, 80, 531

\refindent 
Fabian, A.~C., Rees, M.~J., Stella, L., \& White, N.~E. 1989, \mnras, 238, 729

\refindent 
Ferrarese, L., Ford, H.~C., \& Jaffe, W. 1996, \apj, 470, 444


\refindent 
Forbes, D.~A., Ward, M.~J., DePoy, D.~L., Boisson, C., \& Smith, M.~S. 1992,
\mnras, 254, 509



\refindent 
Gaskell, C.~M., \& Ferland, G.~J. 1984, \pasp, 96, 393

\refindent 
Gonz\'alez-Delgado, R.~M., \& P\'erez, E. 1996, \mnras, 281, 1105

\refindent 
Gordon, K.~D., Calzetti, D., \& Witt, A.~N. 1997, \apj, 487


\refindent 
Heckman, T.~M. 1980, \aa, 87, 152

\refindent 
Ho, L.~C. 1998c, in Observational Evidence for Black Holes in the Universe,
ed. S.~K. Chakrabarti (Dordrecht: Kluwer), 157

\refindent 
Ho, L.~C. 1999a, in  The AGN-Galaxy Connection, ed. H.~R. Schmitt, L.~C. Ho, \&
A.~L. Kinney (Advances in Space Research), in press

\refindent 
Ho, L.~C. 1999b, in preparation

\refindent 
Ho, L.~C., \etal 1999b, in preparation

\refindent 
Ho, L.~C., Filippenko, A.~V., \& Sargent, W.~L.~W. 1995, \apjs, 98, 477

\refindent 
Ho, L.~C., Filippenko, A.~V., \& Sargent, W.~L.~W. 1996, \apj, 462, 183

\refindent 
Ho, L.~C., Filippenko, A.~V., \& Sargent, W.~L.~W. 1997a, \apj, 487, 568

\refindent 
Ho, L.~C., Filippenko, A.~V., \& Sargent, W.~L.~W. 1997b, \apjs, 112, 315


\refindent 
Ho, L.~C., Van Dyk, S.~D., Pooley, G.~G., Sramek, R.~A., \& Weiler,
K.~W. 1999a, \aj, submitted

\refindent 
Hummel, E., van der Hulst, J.~M., \& Dickey, J.~M. 1984, \aa, 134, 207

\refindent 
Impey, C.~D., Wynn-Williams, C.~G., \& Becklin, E.~E. 1986, \apj, 309, 572

\refindent 
Ishisaki, Y., \etal 1996, PASJ, 48, 237


\refindent 
Jones, D.~L., \etal 1986, \apj, 305, 684

\refindent 
Jones, D.~L., Terzian, Y., \& Sramek, R.~A. 1981, \apj, 246, 28

\refindent 
Jones, D.~L., \& Wehrle, A.~E. 1997, \apj, 484, 186

\refindent 
Kellermann, K.~I., Sramek, R.~A., Schmidt, M., Shaffer, D.~B., \& Green,
R.~F. 1989, \aj, 98, 1195

\refindent 
Kormendy, J., \etal 1997, \apj, 473, L91

\refindent 
Laor, A., \& Draine, B. 1993, \apj, 402, 441

\refindent 
Lasota, J.-P., Abramowicz, M.~A., Chen, X., Krolik, J., Narayan, R., \& Yi, I.
 1996, \apj, 462, 142


\refindent 
Maiolino, R., Ruiz, M., Rieke, G.~H., \& Keller, L.~D. 1995, \apj, 446, 561

\refindent 
Malkan, M.~A., \& Oke, J.~B. 1983, \apj, 265, 92

\refindent 
Malkan, M.~A., \& Sargent, W.~L.~W. 1982, \apj, 254, 22 

\refindent 
Maoz, D., Filippenko, A.~V., Ho, L.~C., Rix, H.-W., Bahcall, J.~N.,
Schneider, D.~P., \& Macchetto, F.~D. 1995, \apj, 440, 91

\refindent 
Maoz, D., Koratkar, A.~P., Shields, J.~C., Ho, L.~C., Filippenko, A.~V., \&
Sternberg, A. 1998, \aj, 116, 55

\refindent
Mathis, J.~S. 1994, \apj, 422, 176

\refindent
McKee, C.~F., \& Petrosian, V. 1974, \apj, 189, 17

\refindent 
Murphy, E.~M., Lockman, F.~J., Laor, A., \& Elvis, M. 1996, ApJS, 105, 369

\refindent 
Mushotzky, R.~F., \& Wandel, A. 1989, \apj, 339, 674

\refindent 
Nandra, K., George, I.~M., Mushotzky, R.~F., Turner, T.~J., \& Yaqoob, T.
1997, \apj, 477, 602

\refindent 
Neugebauer, G., \etal 1980, \apj, 238, 502

\refindent 
Nicholson, K.~L., Reichert, G.~A., Mason, K.~O., Puchnarewicz, E.~M.,
Ho, L.~C., Shields, J.~C., \& Filippenko, A.~V. 1998, \mnras, 300, 893

\refindent 
Pauliny-Toth, I.~I.~K., Preuss, E., Witzel, A., Graham, D., Kellermann,
K.~I., Ronnang, B. 1981, \aj, 86, 371

\refindent 
Pei, Y.~C. 1992, \apj, 395, 130


\refindent 
Reichert, G.~A., Mushotzky, R.~F., Petre, R., \& Holt, S.~S. 1985, \apj, 296,
69

\refindent 
Reid, M.~J., Biretta, J.~A., Junor, W., Muxlow, T.~W.~B., \& Spencer,
R.~E. 1989, \apj, 336, 112

\refindent 
Reynolds, C.~S., Di Matteo, T., Fabian, A.~C., Hwang, U., \& Canizares, C.~R.
1996, \mnras, 283, L111

\refindent 
Rieke, G.~H., \& Lebofsky, M.~J. 1978, \apj, 220, L37

\refindent 
Sadler, E.~M., Slee, O.~B., Reynolds, J.~E., \& Roy, A.~L. 1995, \mnras, 276,
1373
 
\refindent 
Sanders, D.~B., Phinney, E.~S., Neugebauer, G., Soifer, B.~T., \& Matthews, K.
 1989, \apj, 347, 29


\refindent 
Schilizzi, R.~T. 1976, \aj, 81, 946

\refindent 
Serlemitsos, P., Ptak, A., \& Yaqoob, T. 1996, in The Physics of LINERs in 
View of Recent Observations, ed. M. Eracleous et al.  (San Francisco: ASP), 70

\refindent 
Shields, G.~A. 1978, \nat, 272, 706

\refindent 
Shuder, J.~M., \& Osterbrock, D.~E. 1981, \apj, 250, 55

\refindent 
Soltan, A. 1982, \mnras, 200, 115
 
\refindent 
Spencer, R.~E., \& Junor, W. 1986, \nat, 321, 753

\refindent 
Stiavelli, M., Peletier, R.~F., \& Carollo, C.~M. 1997, \mnras, 285, 181

\refindent 
Terashima, Y., Kunieda, H., Misaki, K., Mushotzky, R.~F., Ptak, A.~F., \&
Reichert, G.~A. 1998, \apj, 503, 212

\refindent 
Tripp, T.~M., Bechtold, J., Green, R.~F. 1994, \apj, 433, 533

\refindent 
Tsvetanov, Z.~I., Hartig, G.~F., Ford, H.~C., Kriss, G.~A., Dopita, M.~A.,
Dressel, L.~L., \& Harms, R.~J. 1998, in  Proceedings of the M87 Workshop
(Lecture Notes in Physics: Springer Verlag), in press

\refindent 
Turner, J.~L., \& Ho, P.~T.~P. 1994, \apj, 421, 122

\refindent 
Turner, T.~J., George, I.~M., Nandra, K., \& Mushotzky, R.~F. 1997, \apjs, 113, 23

\refindent 
Turner, T.~J., \& Pounds, K.~A. 1989, \mnras, 240, 833

\refindent 
Voit, G.~M. 1991, \apj, 379, 122

\refindent 
Willner, S.~P., Elvis, M., Fabbiano, G., Lawrence, A., \& Ward, M.~J. 1985,
\apj, 299, 443

\refindent 
Worral, D.~M., \& Birkinshaw, M. 1994, \apj, 427, 134

\refindent 
Zirbel, E.~L., \& Baum, S.~A. 1998, \apjs, 114, 177

\clearpage
\begin{figure}
\plotone{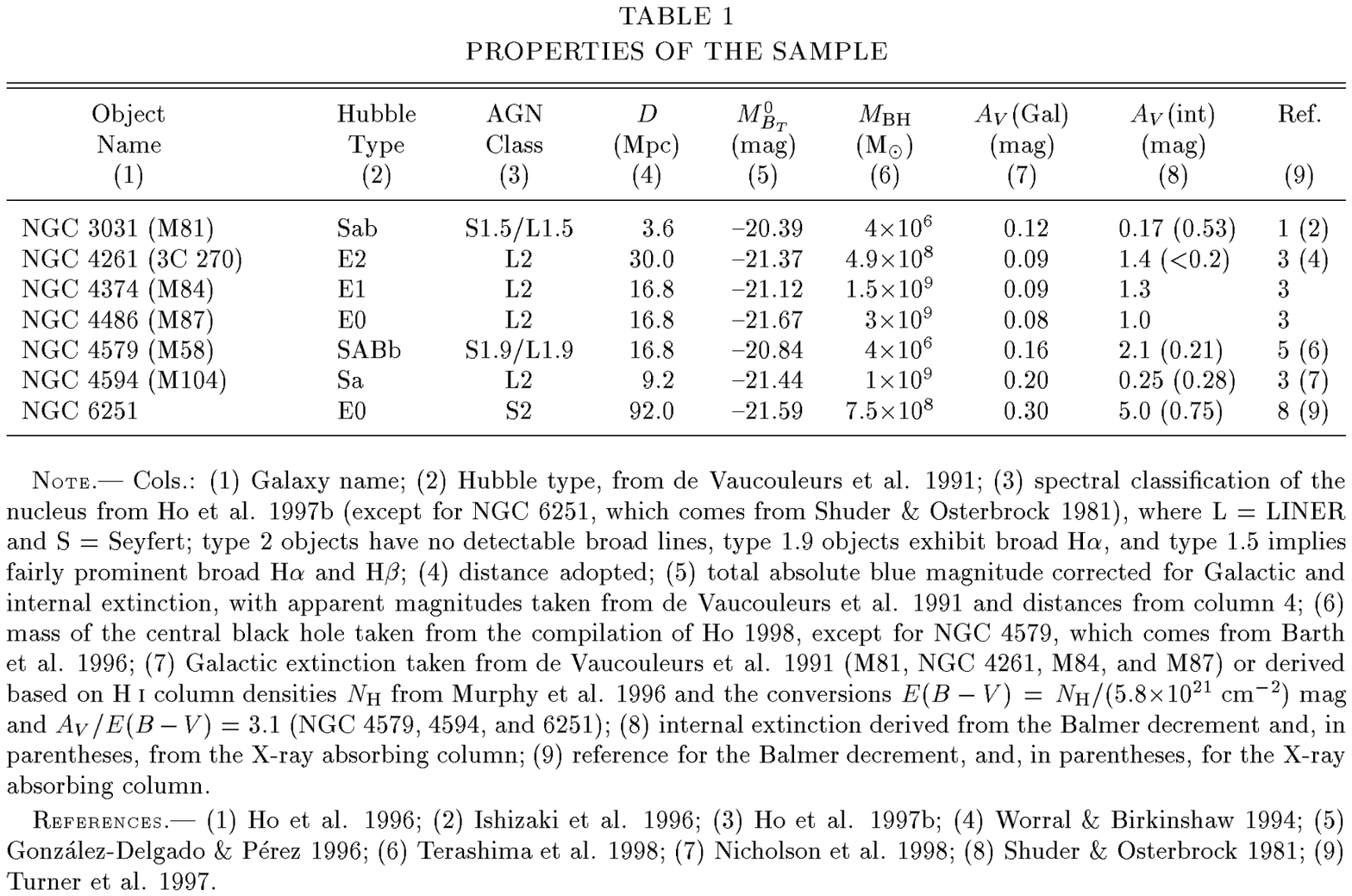}
\end{figure}

\clearpage
\begin{figure}
\plotone{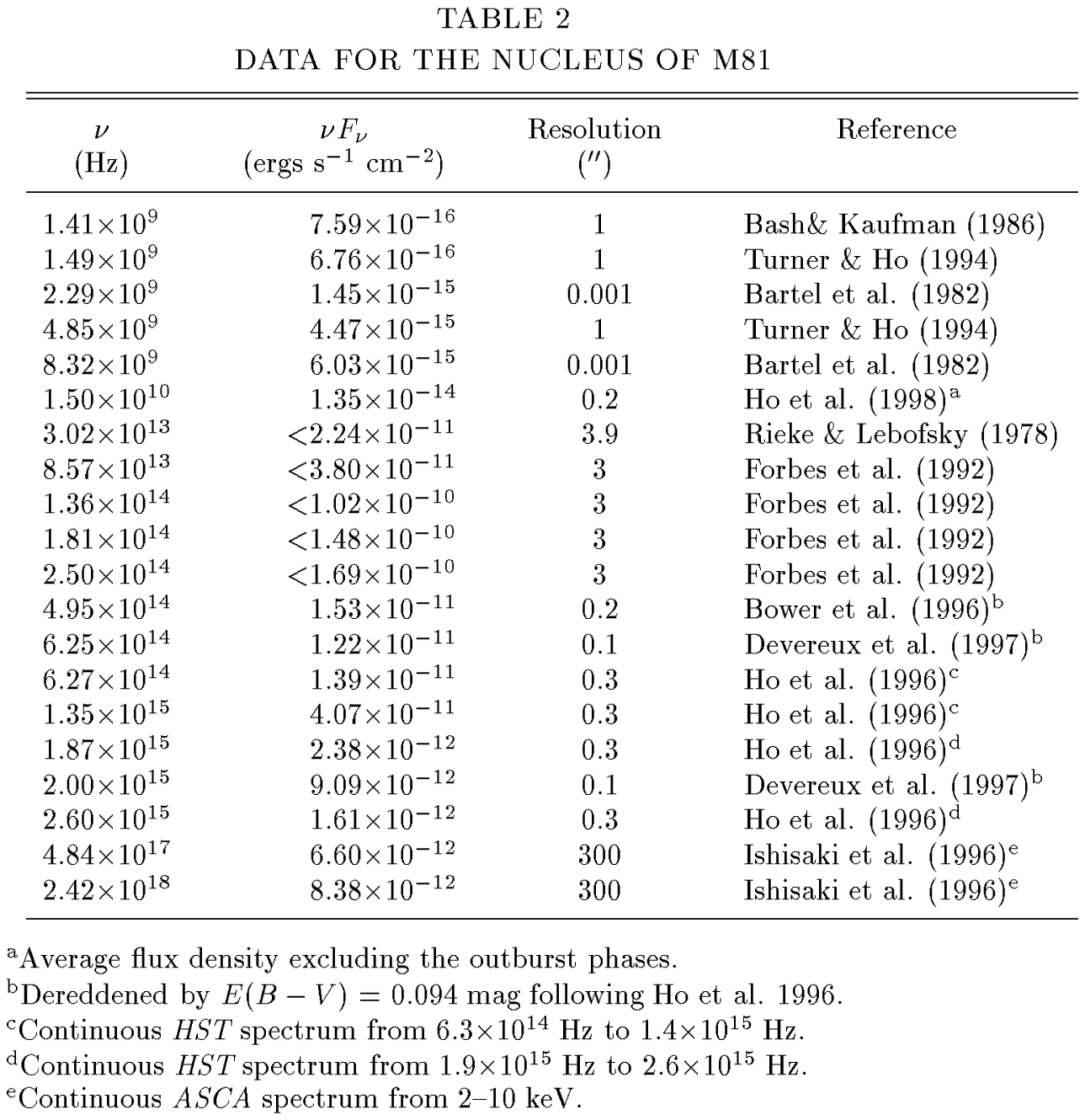}
\end{figure}

\clearpage
\begin{figure}
\plotone{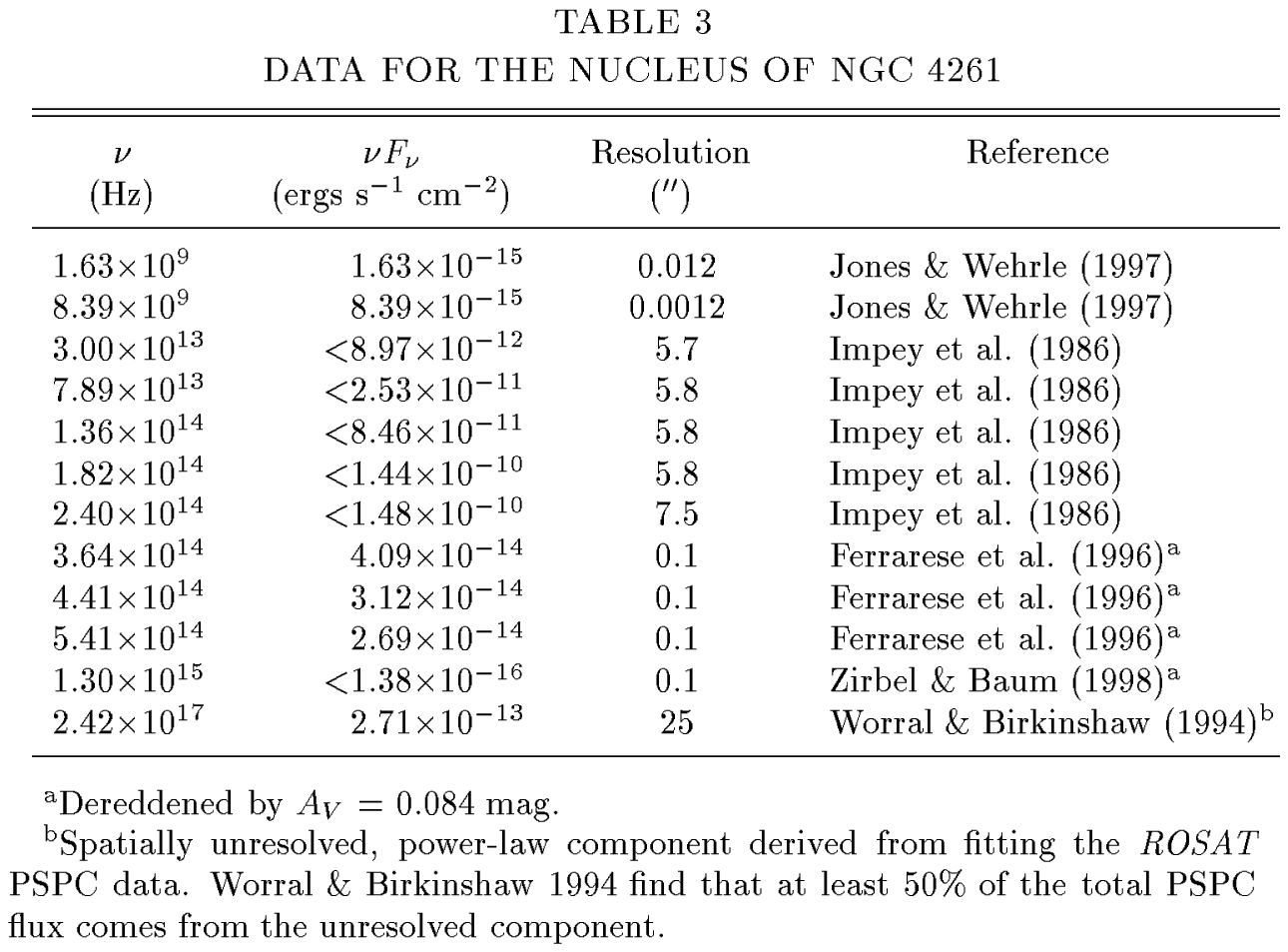}
\end{figure}

\clearpage
\begin{figure}
\plotone{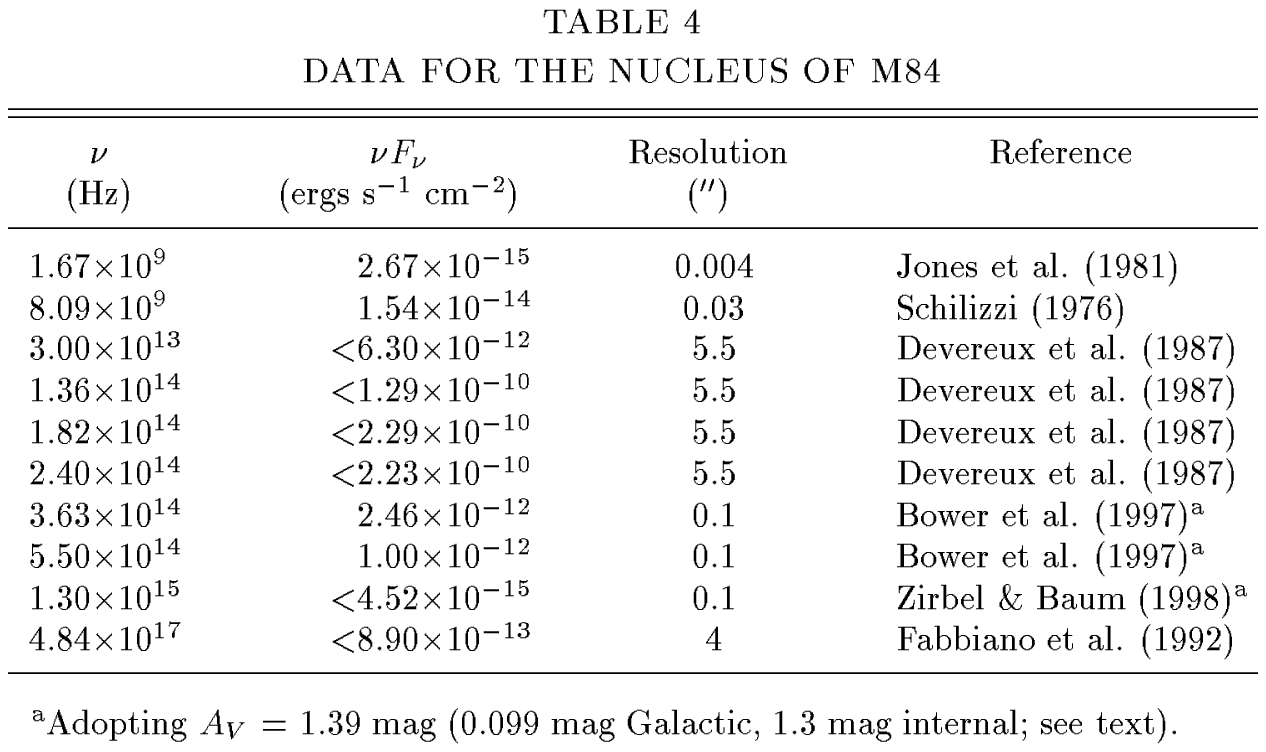}
\end{figure}

\clearpage
\begin{figure}
\plotone{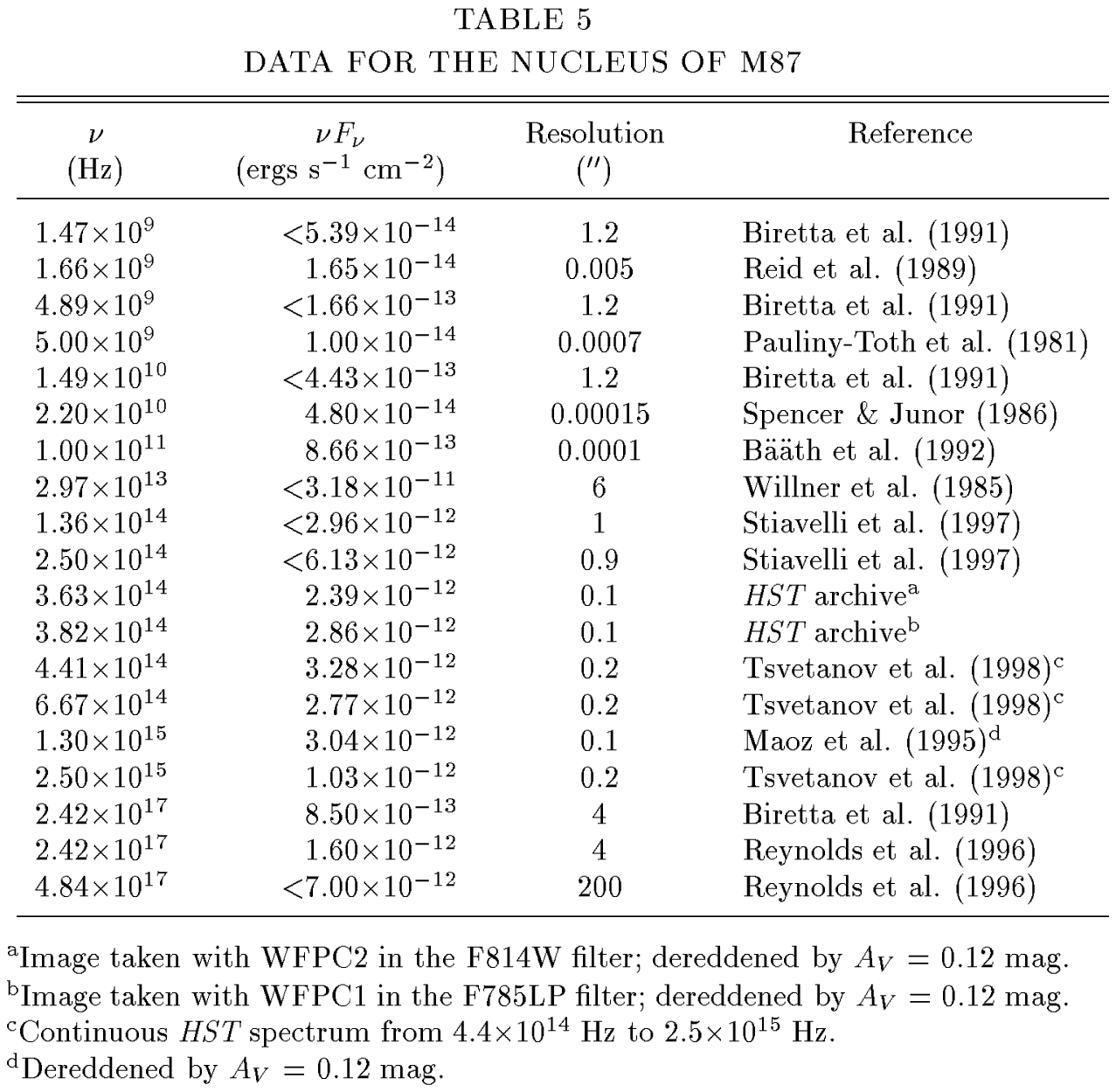}
\end{figure}

\clearpage
\begin{figure}
\plotone{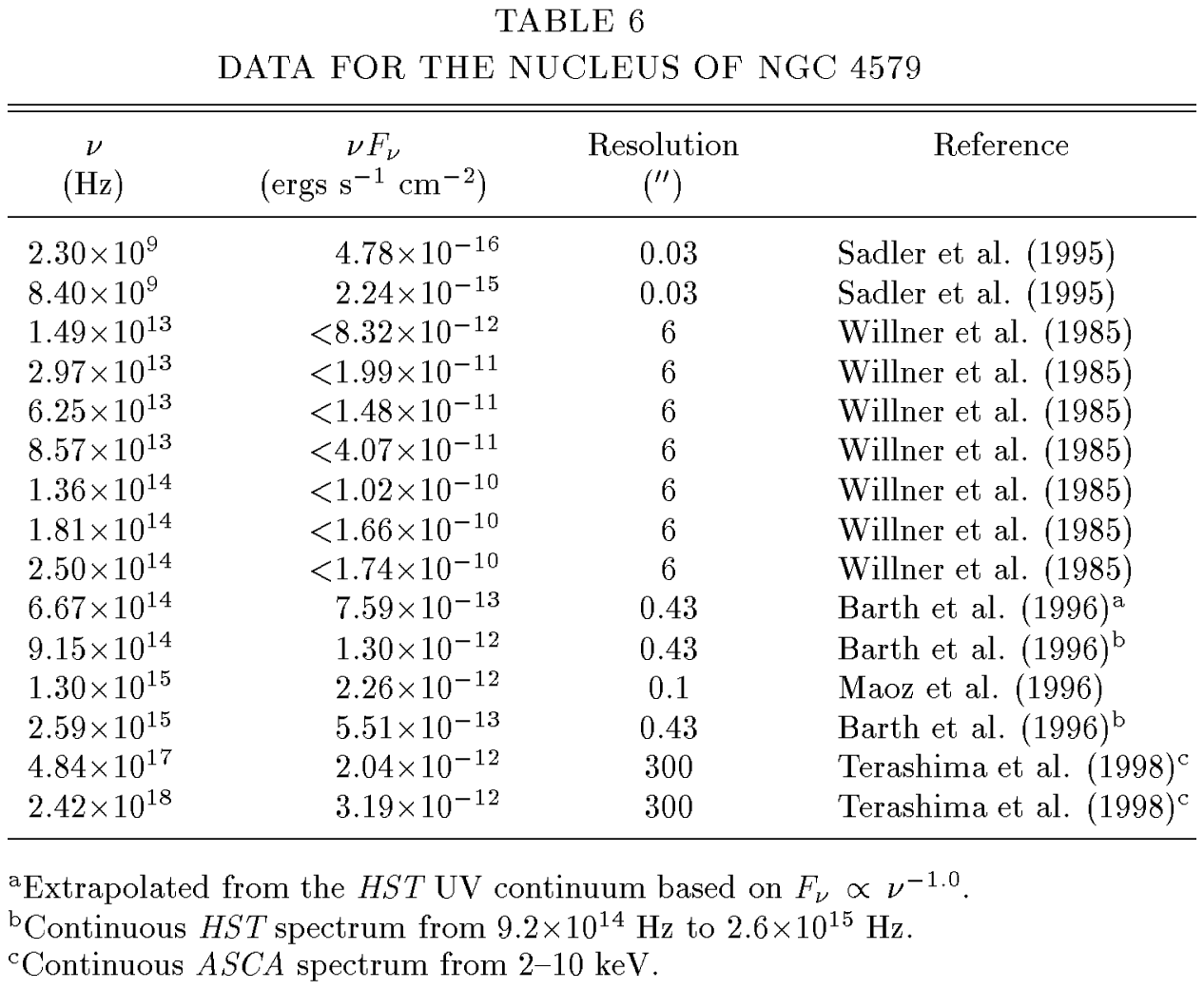}
\end{figure}

\clearpage
\begin{figure}
\plotone{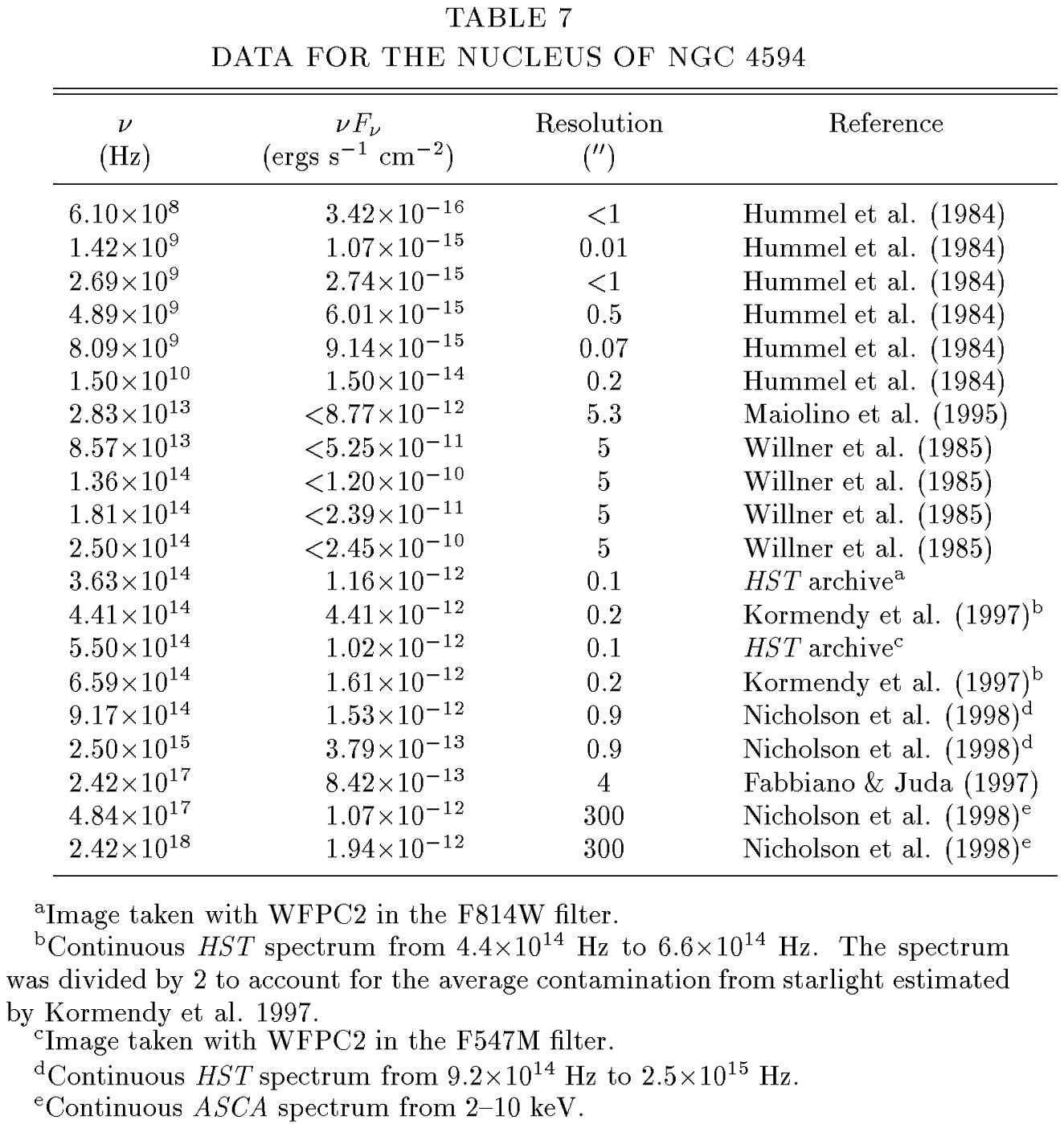}
\end{figure}

\clearpage
\begin{figure}
\plotone{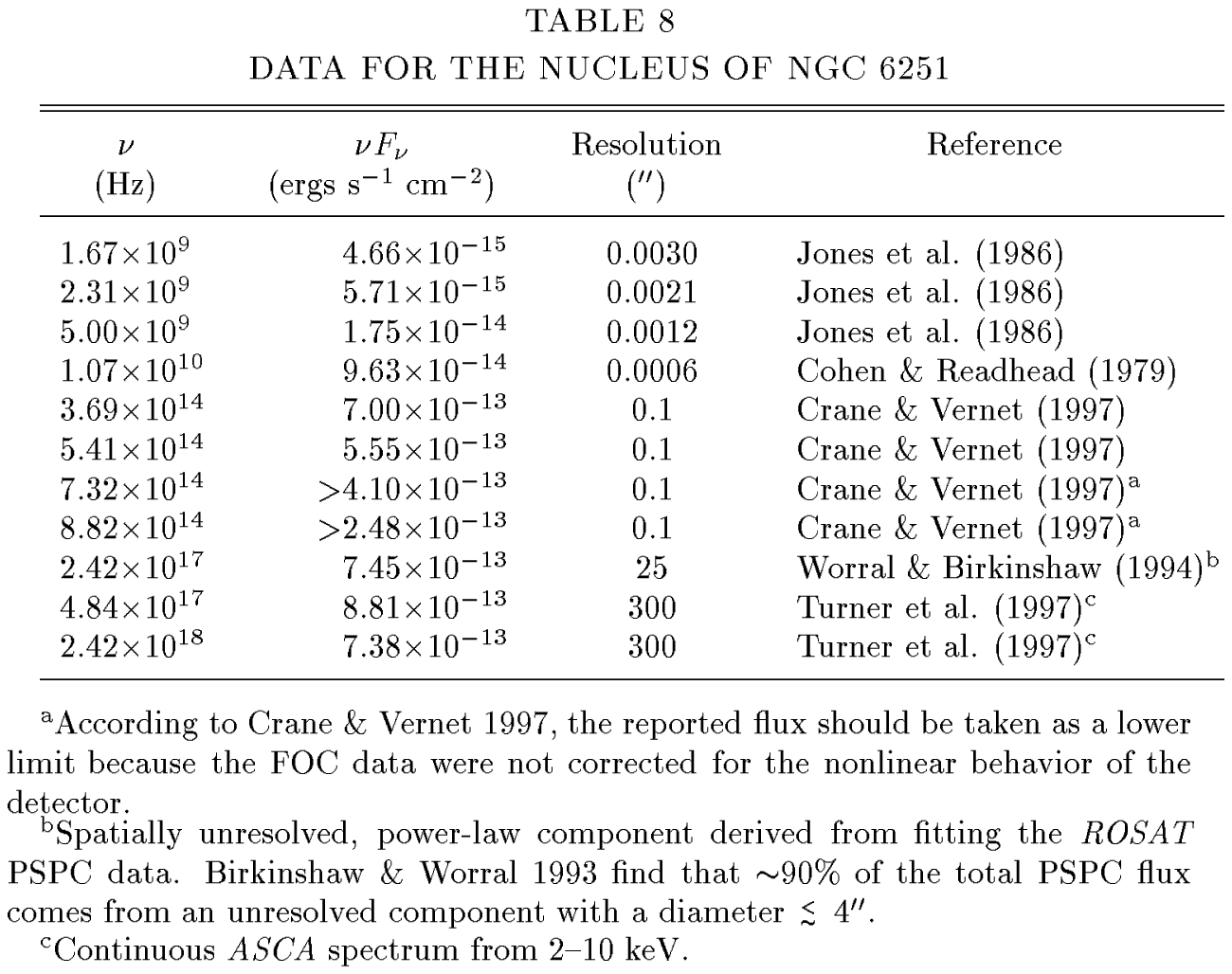}
\end{figure}

\clearpage
\begin{figure}
\plotone{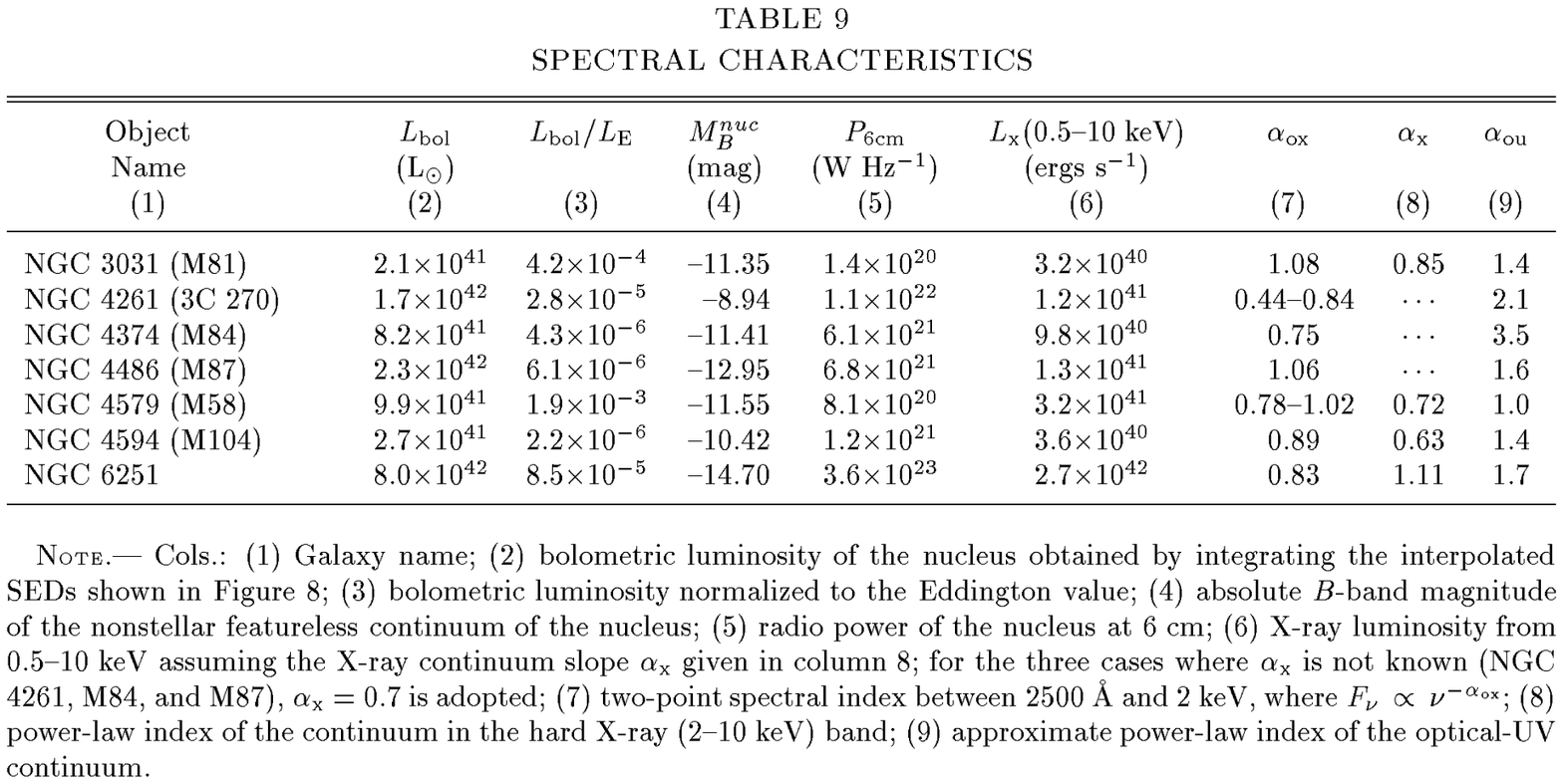}
\end{figure}

%
\clearpage
\begin{figure}
\plotone{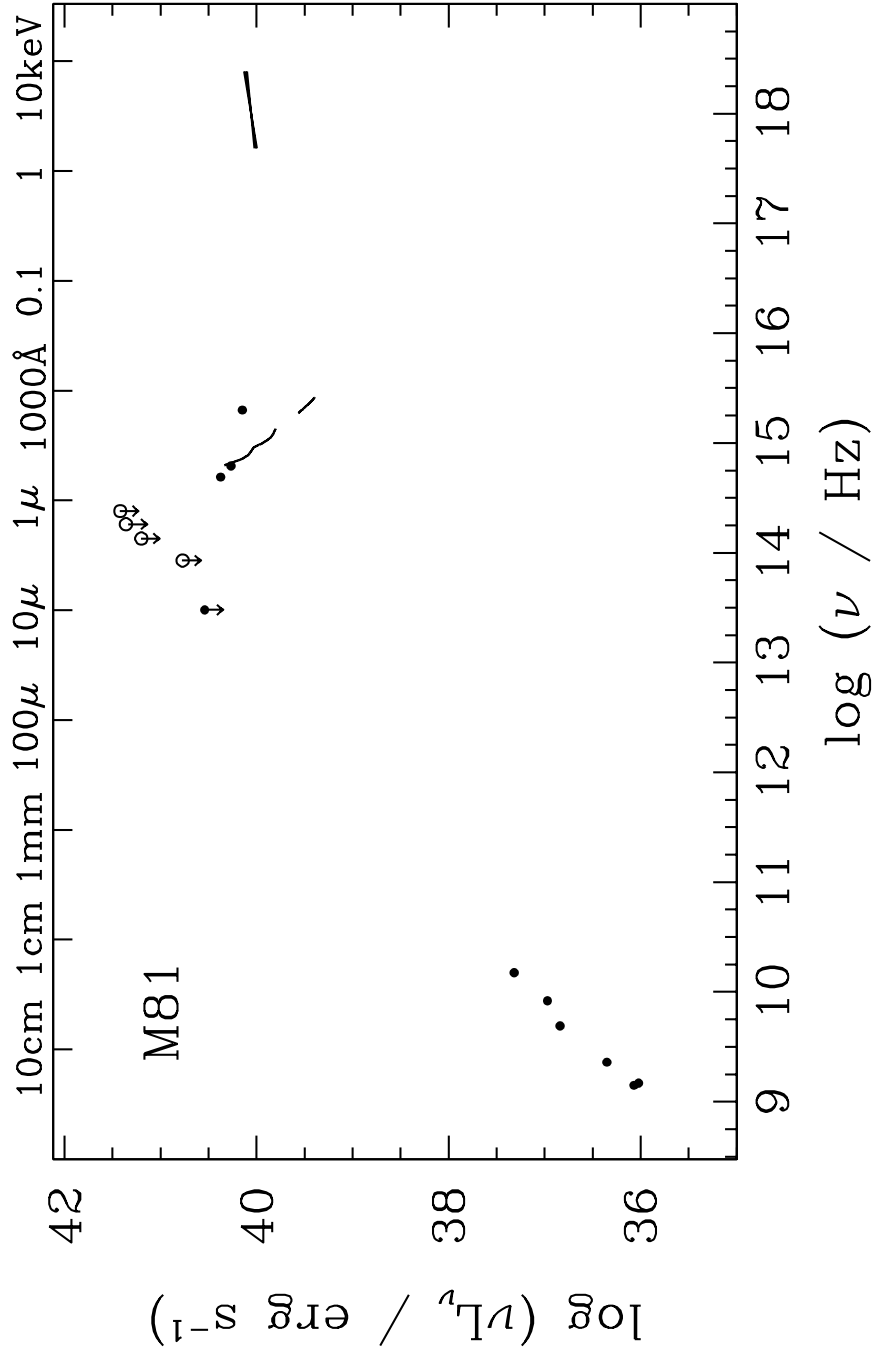}
\caption{
Spectral energy distribution of M81 (see Table 2).  In this and in in 
subsequent figures, open symbols denote data that are believed to be 
heavily contaminated by non-nuclear emission.
}
\end{figure}

\clearpage
\begin{figure}
\plotone{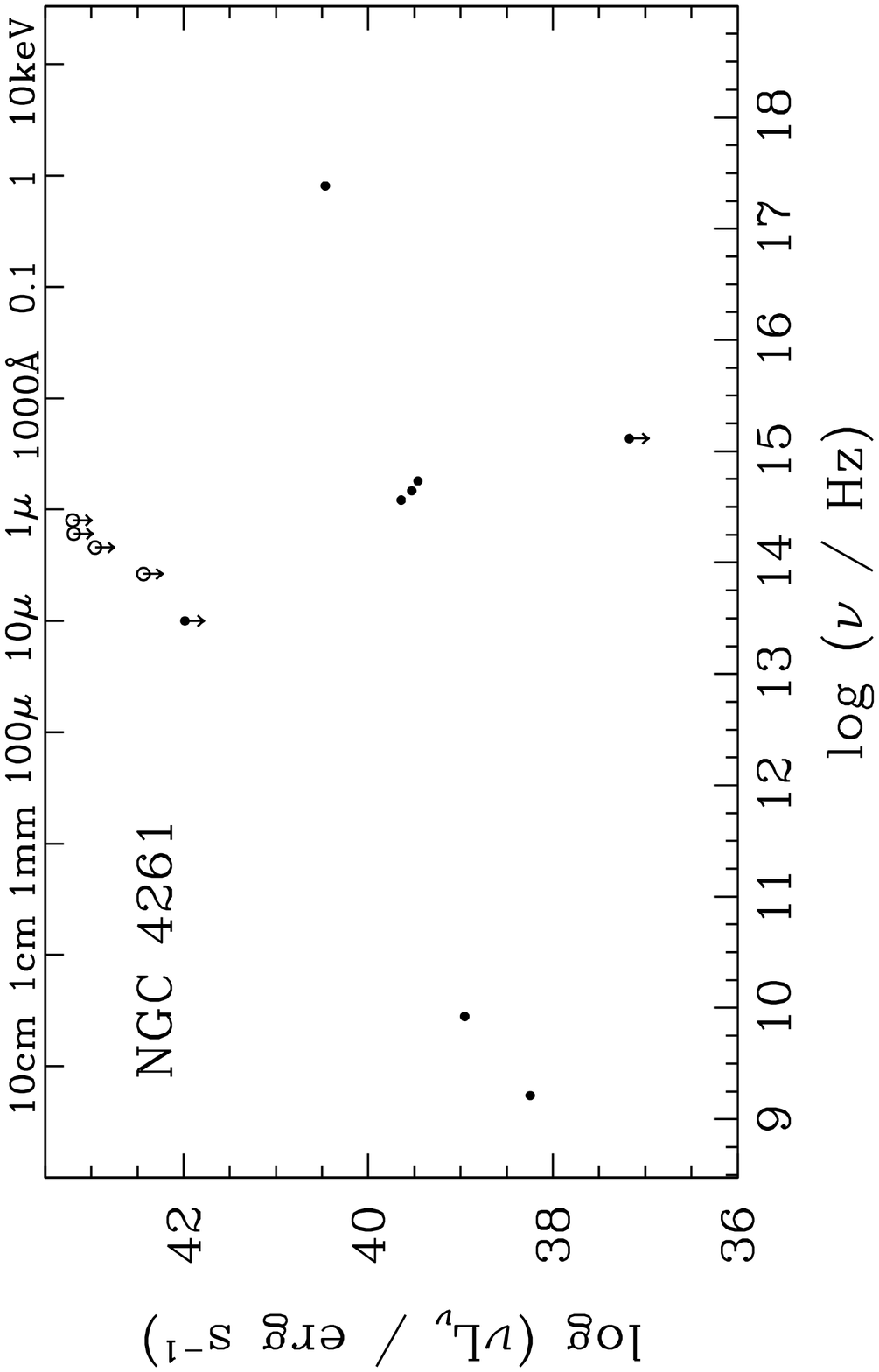}
\caption{
Spectral energy distribution of NGC 4261 (see Table 3).
}
\end{figure}

\clearpage
\begin{figure}
\plotone{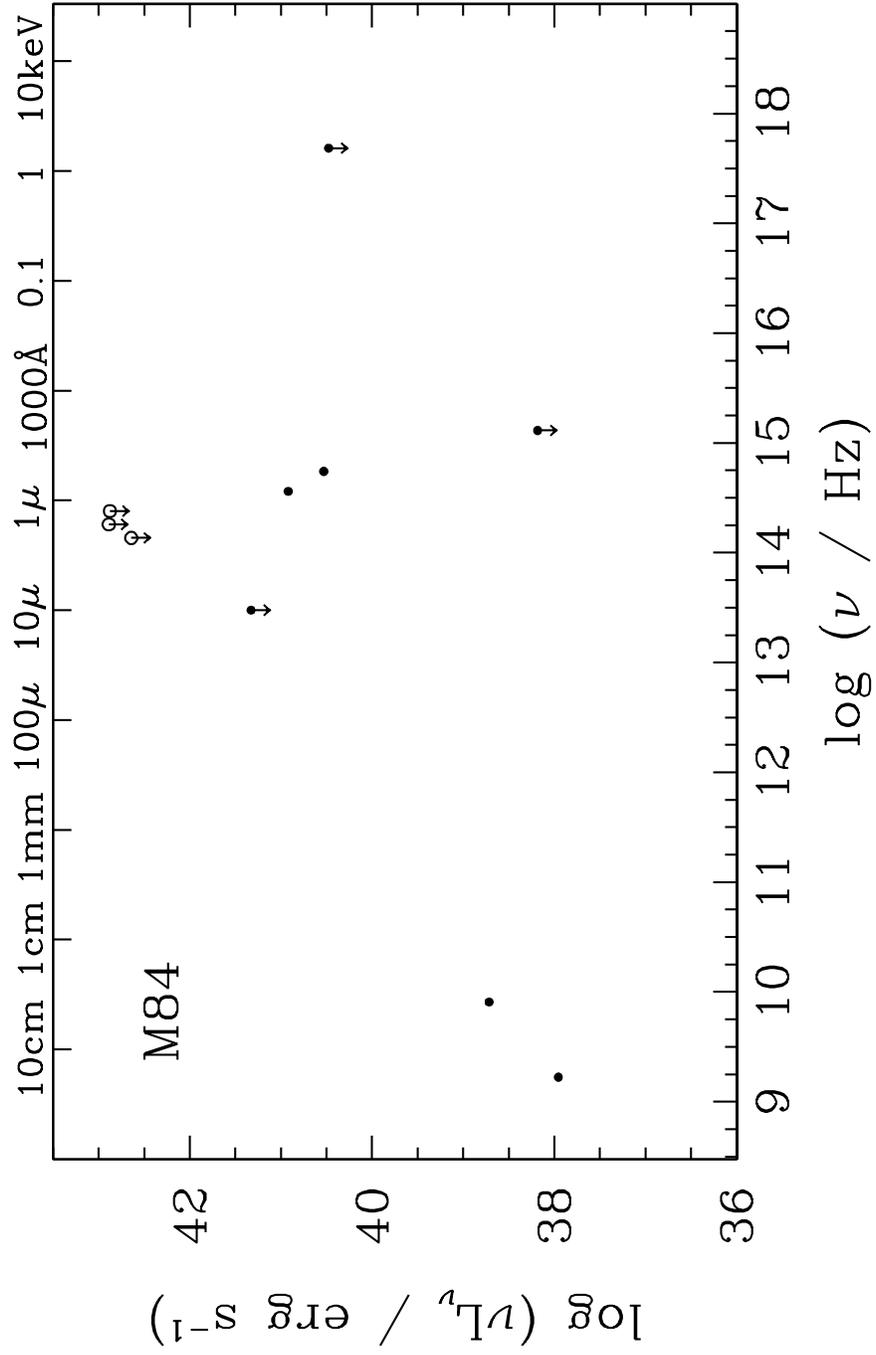}
\caption{
Spectral energy distribution of M84 (see Table 4).
}
\end{figure}

\clearpage
\begin{figure}
\plotone{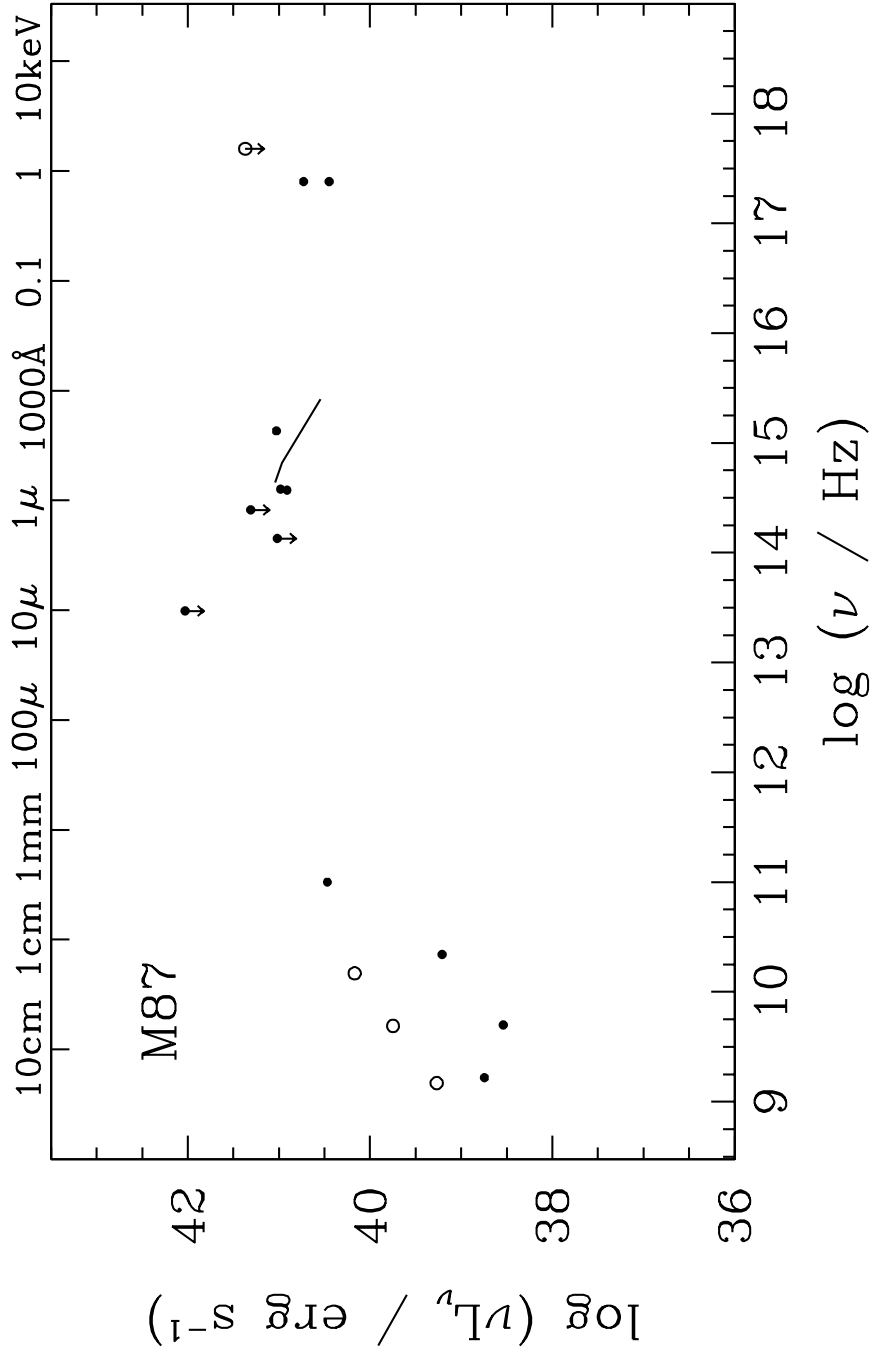}
\caption{
Spectral energy distribution of M87 (see Table 5).  The three radio points 
in {\it open} symbols were obtained with a relatively large beam.  
They are plotted along with the VLBI points ({\it solid} symbols) to illustrate
the effects of contamination by the radio jet.
}
\end{figure}

\clearpage
\begin{figure}
\plotone{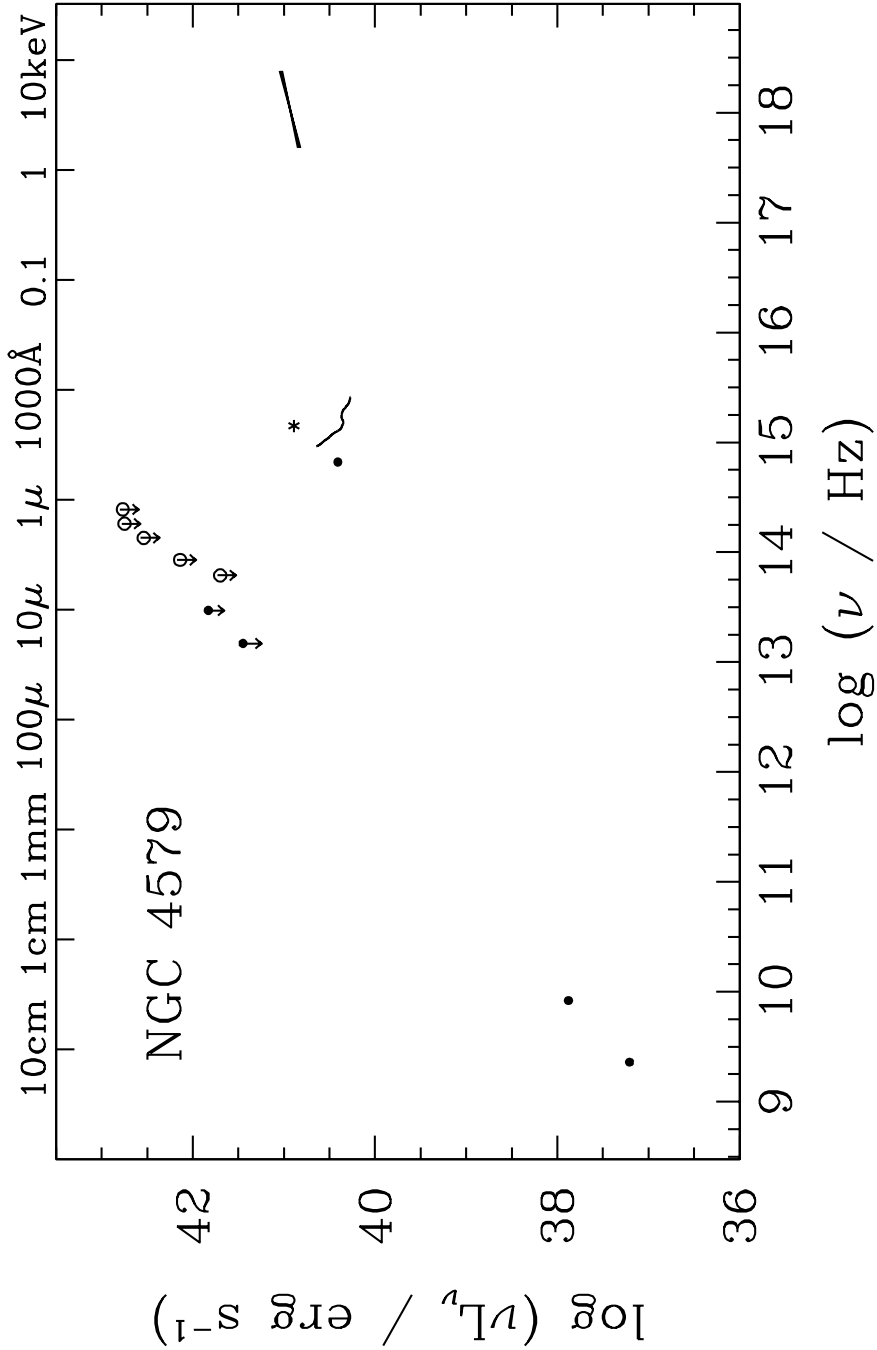}
\caption{
Spectral energy distribution of NGC 4579 (see Table 6).  The UV point 
(log~$\nu$ = 15.1) measured with the FOC ({\it asterisk}) is a factor of 
3--4 higher than the corresponding flux in the FOS spectrum ({\it solid 
line}).  Barth \etal (1996) and Maoz \etal (1998) argue that the difference 
is due to variability.
}
\end{figure}

\clearpage
\begin{figure}
\plotone{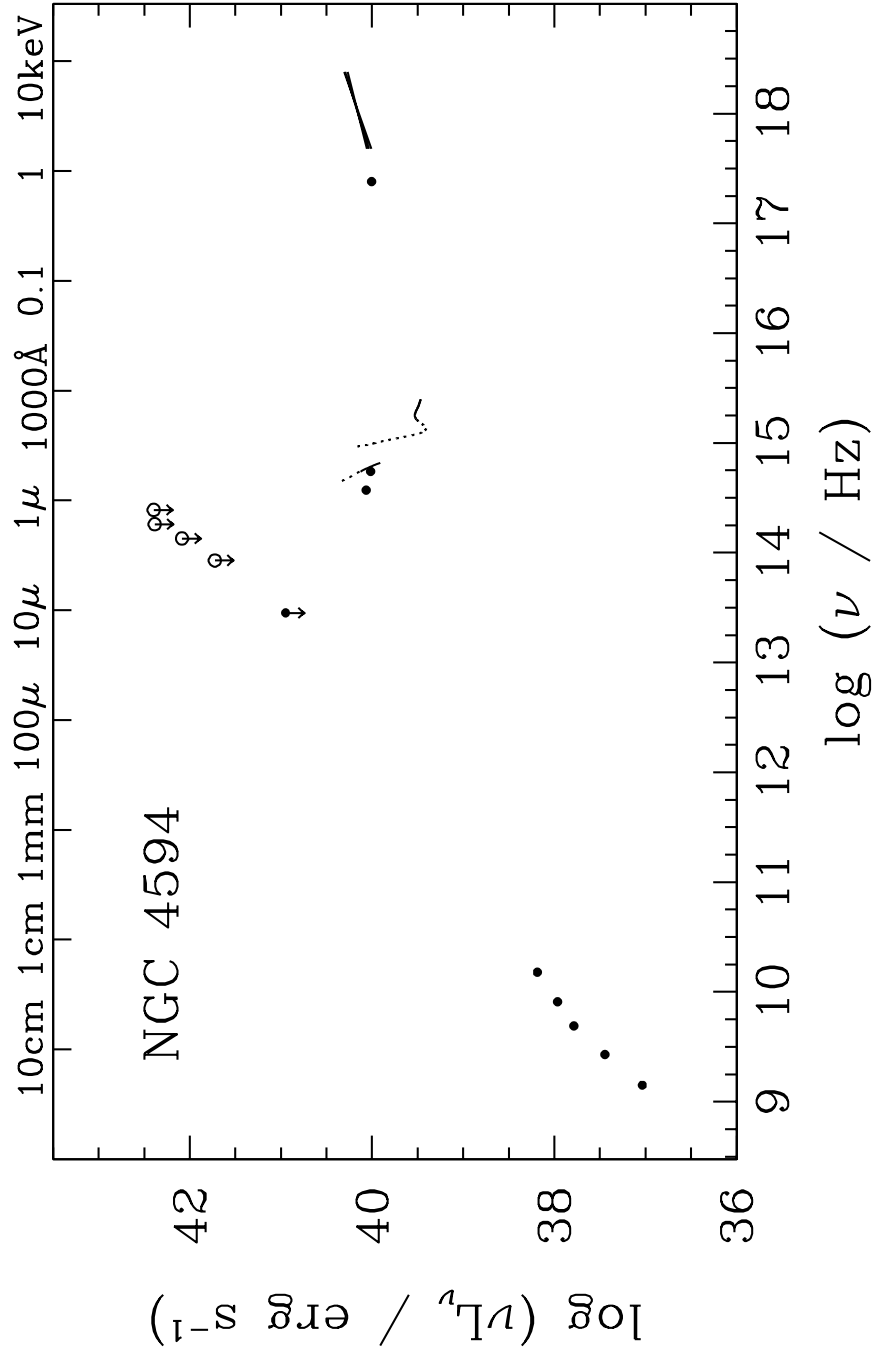}
\caption{
Spectral energy distribution of NGC 4594 (see Table 7).  The portions of the 
FOS spectra that are contaminated by starlight are plotted as {\it dotted 
lines}.
}
\end{figure}

\clearpage
\begin{figure}
\plotone{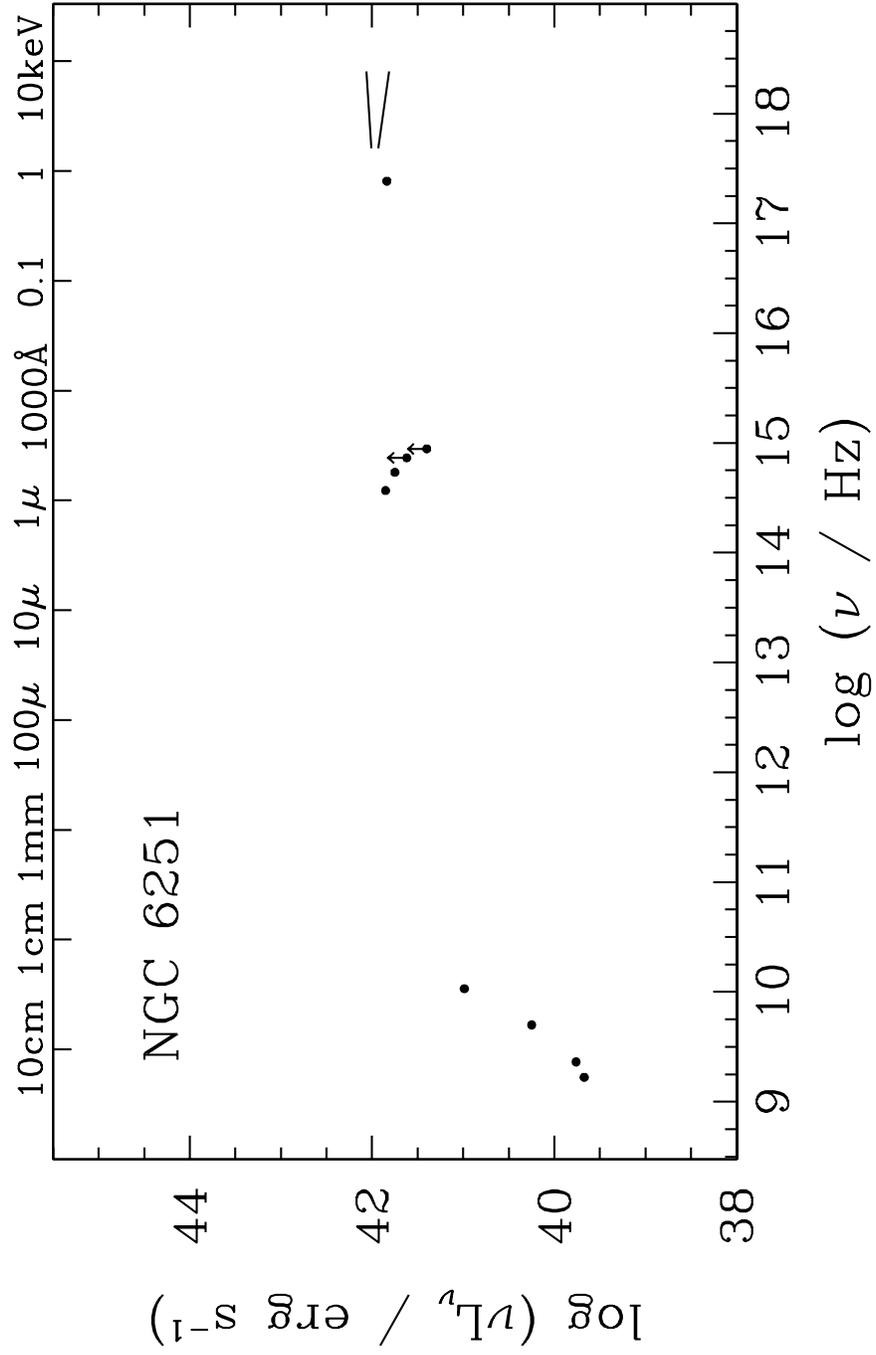}
\caption{
Spectral energy distribution of NGC 6251 (see Table 8).
}
\end{figure}

\clearpage
\begin{figure}
\plotone{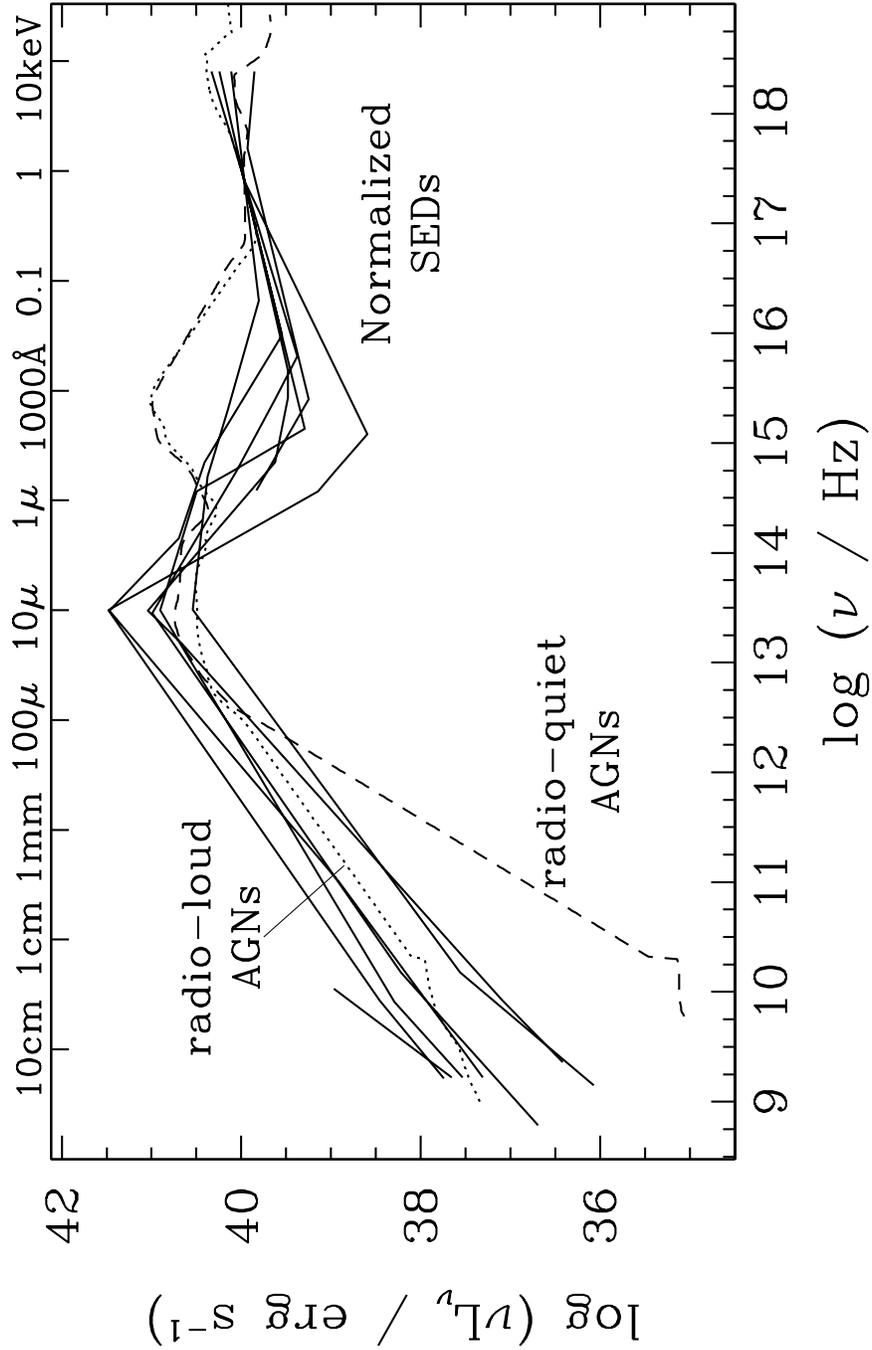}
\caption{
Interpolated SEDs of all the objects ({\it solid lines}) normalized to the 
1 keV luminosity of M81.  The median radio-loud ({\it dotted line}) and 
radio-quiet ({\it dashed line}) SEDs of Elvis \etal (1994), normalized 
similarly, are overplotted for comparison.
}
\end{figure}

\clearpage
\begin{figure}
\plotone{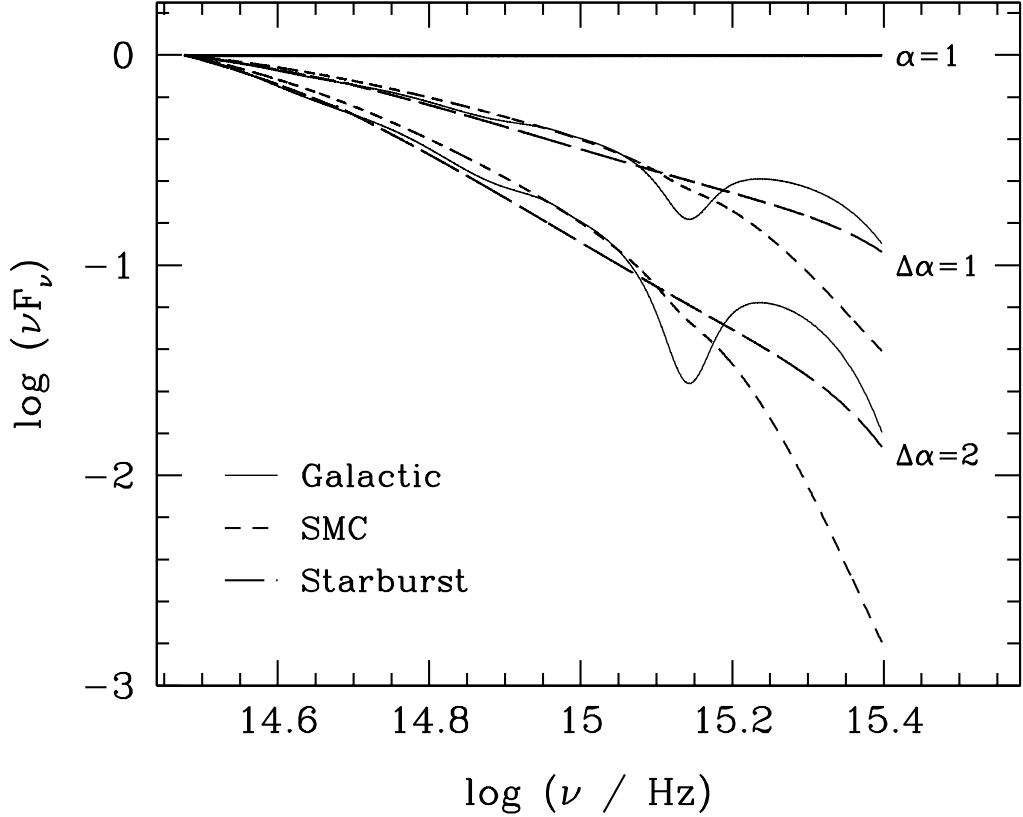}
\caption{
The effect of different extinction laws on the SED from 1200 \AA\ to
1 \micron.  The intrinsic (unextincted) continuum is assumed to be a
single power law with $\alpha$ = 1 (top curve, {\it heavy solid} line),
and it was reddened such that the optical-UV slope between 6500 \AA\
and 2500 \AA\ changes by $\Delta\alpha$ = 1 (middle curves) and
$\Delta\alpha$ = 2 (bottom curves).  The SEDs have been normalized at 1
\micron\ on an arbitrary flux scale. The extinction curves used are those of
the Galaxy (Cardelli \etal 1989; {\it light solid} line), the SMC
(Bouchet \etal 1985; {\it short dashed} line), and the empirical curve 
 for starburst galaxies (Calzetti \etal 1994; {\it long dashed} line).
}
\end{figure}

\end{document}